\documentclass[aps,prl,twocolumn,showpacs,amsmath,amssymb,preprintnumbers,10pt,footinbib]{revtex4-1}
\usepackage{epsfig}
\usepackage{amsfonts}
\usepackage{amssymb}
\usepackage{enumitem}
\usepackage{amsthm}
\usepackage{subfig}
\usepackage{extarrows}
\usepackage{bm}
\usepackage{hyperref}
\usepackage{mathrsfs}
\usepackage{bbm}
\usepackage{cancel}
\usepackage[dvipsnames]{xcolor}
\usepackage{accents}
\usepackage{relsize}
\usepackage{float, slashed, graphicx, amssymb, amsmath}
\usepackage{shuffle}
\usepackage{tikz}
\usetikzlibrary{hobby,calc,shapes, positioning, backgrounds,trees, decorations, decorations.markings, decorations.pathmorphing, arrows, shapes.geometric, snakes}
\usepackage{tikz-feynman}
\tikzfeynmanset{compat=1.1.0}
\tikzfeynmanset{
graviton/.style={circle, draw=green!60, fill=green!5, very thick, minimum size=7mm}
}
\tikzfeynmanset{
codot/.style={/tikz/shape=circle,
/tikz/fill=red,/tikz/minimum size=0.1cm,/tikz/inner sep=1.8pt}
}
\tikzfeynmanset{sb/.style=
{/tikz/shape=circle,
/tikz/fill=black,
 /tikz/minimum size=0.3cm,/tikz/inner sep=1.pt
 } }
\tikzfeynmanset{myblob/.style=
{/tikz/shape=ellipse,
/tikz/fill=red,
 /tikz/minimum width=0.5cm,
 } }
 \tikzfeynmanset{myblobFF/.style=
{/tikz/shape=circle,
/tikz/fill=black,
 /tikz/minimum width=0.25cm,
 } }
  \tikzfeynmanset{myblob2/.style=
{/tikz/shape=rectangle,
/tikz/fill=red,
 /tikz/minimum width=0.3cm,
 /tikz/minimum height=0.3cm,
 /tikz/inner sep=1.8pt
 } }
  \tikzfeynmanset{fp/.style=
{/tikz/shape=rectangle,
/tikz/fill=red,
 /tikz/minimum width=0.1cm,/tikz/inner sep=1.8pt
 } }
  \tikzfeynmanset{myvertex/.style=
{/tikz/shape=circle,
/tikz/fill=black,
 /tikz/minimum width=0.05cm,
 } }
  \tikzfeynmanset{HV/.style=
{/tikz/shape=circle,
/tikz/fill={rgb:black,1;white,2},
 /tikz/minimum size=0.01cm,/tikz/inner sep=1.8pt
 } }
\tikzset{box/.pic={\filldraw[fill=black]  (0,0) circle (2.5pt);
				   \filldraw [fill=black] (0.5,0) circle (2.5pt);
			       \draw [line width=5pt] (0,0) -- (0.5,0);}}

\tikzset{wiggle/.style={decorate, decoration=snake}}

 \tikzfeynmanset{HV/.style=
{/tikz/shape=circle,
/tikz/fill={rgb:black,1;white,2},
 /tikz/minimum size=0.01cm,/tikz/inner sep=1.8pt
 } }

\usepackage[T1]{fontenc} 

\newcommand{\tr}{\text{tr}}

\def\sc#1{\overline{#1}}
\makeatletter
\newcommand \UPlus {\mathop {\operator@font \uplus }\limits }
\makeatother
\makeatletter
\newcommand \Bigcup {\mathop {\operator@font \bigcup }\limits }
\makeatother
  \def\LabelNote#1{}
 \def\Label#1{\label{#1}%
  \smash{\hbox to\phipt{\raise1ex\hbox{\tiny[#1]}\hss}}}

\def\nn{\nonumber}

\newcommand{\red}{\color{red}}

\newcommand{\blue}{\color{blue}}

\def\spa#1.#2{\left\langle#1\,#2\right\rangle}
\def\spb#1.#2{\left[#1\,#2\right]}

\def\be{\begin{equation}}
\def\ee{\end{equation}}
\def\bea{\begin{eqnarray}}
\def\eea{\end{eqnarray}}  
\allowdisplaybreaks

\newcommand{\npre}{\mathcal{N}}

\newcommand{\la}{\langle}
\newcommand{\ra}{\rangle}
\newcommand{\mdot}{{\cdot}}

\def\veps{\varepsilon}
 
\begin{document}

\preprint{
	QMUL-PH-24-04
}

\title{Kinematic  Hopf algebra and BCJ numerators at finite $\alpha'$}

\author{Gang Chen$^1$}
\email{gang.chen@nbi.ku.dk}
\author{Laurentiu Rodina$^2$}
\email{laurentiu.rodina@gmail.com}
\author{Congkao Wen$^3$}
\email{c.wen@qmul.ac.uk}

\affiliation{$\mbox{}^{1}$Niels Bohr International Academy,
Niels Bohr Institute, University of Copenhagen,\\
Blegdamsvej 17, DK-2100 Copenhagen \O, Denmark }
\affiliation{$\mbox{}^{2}$Beijing Institute of Mathematical Sciences and Applications, Beijing, 101408, China}
\affiliation{$\mbox{}^{3}$Centre for Theoretical Physics, Department of Physics and Astronomy, 
Queen Mary University of London,\\ Mile End Road, London E1 4NS, United Kingdom}

\begin{abstract}
In this letter, starting from a kinematic Hopf algebra, we first construct a closed-form formula for all Bern-Carrasco-Johansson (BCJ) numerators in Yang-Mills (YM) theory with infinite orders of $\alpha'$ corrections, known as $\rm DF^2+YM$ theory, when coupled to two heavy particles which can be removed through a simple factorization limit. The full $\alpha'$ dependence appears simply in massive physical propagator factors, with factorization strongly constraining the construction. The intricate structure induced by the massive poles also naturally leads us to find a novel closed-form and local expression for BCJ numerators in usual pure YM theory, based directly on the kinematic Hopf algebra.

\end{abstract} 


\maketitle

\section{Introduction}
The discovery of colour-kinematic (CK) duality and the double-copy \cite{Bern:2008qj, Bern:2010ue} has revolutionized the computation of amplitudes, particularly in gravity. The duality relies on finding kinematic numerators that satisfy the same algebraic relations as the corresponding colour factors, such as the Jacobi identity. Then the double copy procedure allows the computation of various amplitudes by mixing and matching different types of such numerators. However, the origin of this duality remains deeply mysterious, in particular because the nature of the kinematic algebra \cite{Monteiro:2011pc, Bjerrum-Bohr:2012kaa,Cheung:2016prv,Chen:2019ywi,Mizera:2019blq,Chen:2021chy,Ben-Shahar:2021zww,Monteiro:2022nqt,Monteiro:2022lwm,Lipstein:2023pih,Ferrero:2020vww,Ben-Shahar:2022ixa,Mafra:2022wml,Fu:2018hpu,Ben-Shahar:2021doh,Borsten:2020zgj,Borsten:2020xbt,Bonezzi:2023pox,Borsten:2022vtg,Armstrong-Williams:2024icu,Brown:2023zxm,Fu:2022esi} at the heart of the duality is not yet understood.  Important progress could be accomplished by finding a more systematic, universal, and efficient approach to  
 constructing BCJ numerators, and recent work suggests this may involve utilising a combinatorial algebra perspective. A novel kinematic algebra, known as the quasi-shuffle Hopf algebra \cite{hoffman2000quasi,Blumlein:2003gb,aguiar2010monoidal,hoffman2017quasi,fauvet2017hopf}, has been discovered within the framework of heavy-mass effective field theory (HEFT) \cite{Brandhuber:2021kpo,Brandhuber:2021bsf, Brandhuber:2022enp}. This kinematic algebra has been subsequently extended to amplitudes and form factors, incorporating finite massive scalar and fermionic contributions \cite{Chen:2022nei,Lin:2022jrp,Lin:2023rwe,FermionKiHA}, and also uncovered in a geometrical context \cite{Cao:2022vou}. Recently, the same algebraic structure has been identified in the $\rm \alpha' F^3+\alpha'^2 F^4$ theory \cite{Chen:2023ekh}\footnote{See also \cite{Bonnefoy:2023imz} for a different approach studying the colour-kinematic duality for $\rm \alpha' F^3+\alpha'^2 F^4$ theory.}, which can be viewed as lower-order terms in the $\alpha'$-expansion of $\rm DF^2+YM$ theory \cite{Johansson:2017srf}. 
This theory can be used to construct amplitudes for both  bosonic and heterotic string theory via the double copy procedure \cite{Huang:2016tag, Azevedo:2018dgo}, as a purely field theoretic alternative to the Kawai-Lewellen-Tye (KLT) relations within string theory \cite{Kawai:1985xq}. 

In this letter, to further explore the applicability and universality of this approach, we reveal the presence of the same kinematic Hopf algebra in the complete $\rm DF^2+YM$ theory. This involves incorporating resummed $\alpha'$ contributions, treating them as massive propagators for the massive gluons and tachyons. The crux of constructing the BCJ numerator from the kinematic Hopf algebra lies in identifying an evaluation map \cite{Brandhuber:2021bsf, Chen:2022nei}. This map connects the abstract combinatorial algebra generators to physically meaningful, gauge-invariant functions that exhibit only physical poles. 
We find the evaluation map is fully determined by imposing the relabeling symmetry and consistent factorization conditions on the  massive and massless  gluon  poles. Consequently, we determine the BCJ numerator, applicable for an arbitrary number of external gluons (with two heavy particles, which can be removed through a factorization limit). By expanding in terms of $\alpha'$, we obtain the local BCJ numerators for massless gluon amplitudes. 

Finally, by exploiting the specific form of the mapping rule, we also derive a novel formula directly for  BCJ numerators in pure YM theory, that is compact, manifestly local and relabeling symmetric. This is a significant improvement from the original results in \cite{Brandhuber:2021kpo,Brandhuber:2021bsf}, as considering HEFT numerators and taking a factorization limit is no longer a required intermediate step. Interestingly, this formula shares several similarities with the expression obtained through completely different methods in \cite{Cheung:2021zvb}, such as in the numbers of terms (i.e. twice the Fubini numbers) and the distribution of gluon labels.

\section{Universal Kinematic Hopf algebra}\label{sec:NumInf}
It has been shown that BCJ numerators in a manifest gauge-invariant form originate from the quasi-shuffle Hopf algebra in YM theory coupled with two heavy particles \cite{Brandhuber:2021bsf,Brandhuber:2022enp,Chen:2022nei} as well as theories with higher-derivative corrections \cite{Chen:2023ekh}.  The BCJ numerators
\begin{align}
     \npre([12\ldots n-2], v)
\end{align}
can be naturally written  in the form of a quasi-shuffle product of algebraic generator $T_{(i)}$ 
\begin{align} \label{eq:Nhat}
   \langle T_{(1)}\star  T_{(2)}\star \cdots \star T_{(n{-}2)}\rangle \, ,
\end{align}
where $i=1,2,\ldots,n{-}2$ represents the external gluons, $v$ is the velocity of the heavy particles, and $\langle \bullet\rangle $ is the evaluation map from the algebra generators to a manifestly gauge invariant function. More details are presented in Appendix \ref{app1}. Already from the study of lower orders in $\alpha'$-expansion \cite{Chen:2023ekh} we find the BCJ numerators are constrained by the heavy-mass factorization behaviour and relabeling symmetry. 
These properties impose stringent constraints on the evaluation map, which then lead to a recursive form  for the BCJ numerator \cite{Chen:2023ekh}
\begin{align}\label{eq:recJ}
    \npre([1\alpha],& v)=(-1)^{n-3} \, { G_{1\alpha}(v)\over  v\mdot p_{1}}   \\
    &+\sum_{\tau_L\subset \alpha}(-1)^{|\tau_R|} 
 \,  {\npre([1\tau_L], v)  G_{\tau_R}(p_{\Theta(\tau_R)}) \over v\mdot p_{1\tau_L}} \, , \nn
\end{align}
where $\{1\tau_L\}\cup\{\tau_R\}=\{1\alpha\}\equiv\{12\ldots n-2\}$, $|\tau_R|$ denotes the number of gluons in $\tau_R$ and $[\bullet]$ denotes the left nested commutator, e.g. $[123]\equiv 123-213-312+321$; and $
    \Theta(\tau_i)= (1\tau_L) \cap \{1,\ldots, \tau_{i[1]}\}$ that consists of all indices to the left of $\tau_i$ and smaller than the first index in $\tau_i$, denoted as $\tau_{i[1]}$.
Finally, the $G$ function takes the following form \cite{Chen:2023ekh}  
\begin{align}\label{eq:Gfun}
    G_{\tau}(x)=  x\mdot F_{\tau}\mdot v+\!\! \sum_{ \sigma_1 i\sigma_2 j  \sigma_3 =\tau} \! x\mdot F_{\sigma_1}\mdot p_{i} W(i\sigma_2j) p_{j}\mdot F_{\sigma_3}\mdot v \,,
\end{align}
with
\begin{align}\label{eq:W2Wprime}
    W(\sigma)&=\sum_{r=1}^{\lfloor |\sigma|/2\rfloor}\sum_{i_1\rho_1 j_1\sigma_2 i_2\rho_2 j_2\ldots i_r\sigma_r j_r\rho_r=\sigma} \nn\\
    & \Big(\prod_{k=1}^rW'(i_k\rho_k j_k)\Big)\Big(\prod_{k=2}^r p_{j_{k-1}}\mdot F_{\sigma_k}\mdot p_{i_k}\Big) \, ,
\end{align}
where we have further introduced $W'$-function, which contains all the $\alpha'$-dependence and will be discussed in detail in the next section.  Below are a few examples of the $G$-functions,   
\begin{align}
    G_{12}(x)&=x\mdot F_{12}\mdot v +x\mdot p_1 W'(12) p_2\mdot v \, , \\
    G_{123}(x)&=x\mdot F_{123}\mdot v +x\mdot F_1\mdot p_2 W'(23) p_3\mdot v \nn\\
    &+x\mdot p_1 W'(12) p_2\mdot F_3\mdot v+x\mdot p_1 W'(123) p_3\mdot v \, .
\end{align}
The amplitudes obtained via colour-kinematic duality then can be expressed as \cite{Brandhuber:2021kpo,Brandhuber:2021bsf,Brandhuber:2022enp}, 
\begin{align} \label{eq:KLT2}
    A(1\alpha,v)&=\sum_{\beta\in S_{n-3}} \mathrm{m}(1\alpha, 1\beta)\,\npre([1\beta],v)\,, 
\end{align}
where $\mathrm{m}(1\alpha, 1\beta)$ is the KLT propagator matrix \cite{Vaman:2010ez}.

As required by the relabeling symmetry of this BCJ numerator, the $W'$-function satisfies the following relations \cite{Chen:2023ekh}
\begin{align}\label{eq:crossingW}
 W'(\rho_1 i_1\rho_2 i_{n-2}\rho_3)&=(-1)^{|\rho_3|+\delta_{0,|\rho_1|}} W'(i_1[\rho_1] \rho_2 [\rho^{\rm rev}_3]i_{n-2})\, , \nn\\
    W'(\rho_1 i_{n-2}\rho_2 i_{1}\rho_3)&=(-1)^{n-2}W'(\rho_3^{\rm rev} i_1\rho^{\rm rev}_2 i_{n-2}\rho^{\rm rev}_1)\, ,
\end{align}
where $|\rho|$ is the size of $\rho$ and $\rho^{\rm rev}$ denotes the its reverse. 
For example, at $n=8$, we have 
\begin{align}
	W'(231645)&=-W'(1[23][45]6) \, , \\
	W'(126345)&=-W'(12[543]6) \, , 
\end{align}
and an independent basis is given by $W'(1i_2 i_3 i_4 i_56)$.

\section{General properties of the $W'$-function from factorizations}

We will consider the theory of interest, the $\rm DF^2+YM$ theory \cite{Johansson:2017srf}, which has been an interesting model for study colour-kinematics duality with $\alpha'$ corrections (see e.g. \cite{Azevedo:2017lkz,Garozzo:2018uzj,Menezes:2021dyp,Carrasco:2022sck,Carrasco:2022lbm, Carrasco:2023wib, Bonnefoy:2023imz,  Garozzo:2024myw}). The theory contains a  massless gluon, as well as a tachyon and a massive gluon (both with $m^2=-1/\alpha'$), and it has played a vital role in constructing string amplitudes via double copy methods \cite{Huang:2016tag, Johansson:2017srf, Azevedo:2018dgo, Azevedo:2019zbn}. 

In order to determine the corresponding $W'$-function for BCJ numerators in $\rm DF^2+YM$ theory, with two heavy particles, we first determine its general properties that follow from the amplitudes in this theory. We begin by considering the HEFT BCJ numerator $\npre ({[1\ldots (n{-}3) {q}]},v)$ with a cut on a massive propagator
\begin{align}
\begin{tikzpicture}[baseline={([yshift=-1.8ex]current bounding box.center)}]\tikzstyle{every node}=[font=\small]    
   \begin{feynman}
    \vertex (a)[myblob2]{};
    \vertex [above=0.35cm of a](aa){\red{$\bm\times$}};
     \vertex [above=0.7cm of a](b)[dot]{};
     \vertex [left=0.7cm of b](c);
     \vertex [left=0.3cm of b](c23);
     \vertex [above=0.23cm of c23](v23)[dot]{};
    \vertex [above=.5cm of c](j1){$1$};
    \vertex [right=.9cm of j1](j2){$\cdots n{-}3~~~~~$};
    \vertex [right=0.6cm of j2](j3){${q}$};
     \vertex [below=0.3cm of v23](l4){$n{-}2~~~~$};
   	 \diagram*{(a) -- [thick] (b),(b) -- [thick] (j1),(v23) -- [thick] (j2),(b)--[thick](j3)};
    \end{feynman}  
  \end{tikzpicture}\!\!\!\!\xleftrightarrow[]{\text{massive cut}} \,  \, \begin{tikzpicture}[baseline={([yshift=-0.8ex]current bounding box.center)}]\tikzstyle{every node}=[font=\small]    
   \begin{feynman}
    \vertex (a)[]{$(n{-}1)_{g'}$};
      \vertex [above=0.9cm of a](b)[dot]{};
     \vertex [left=0.7cm of b](c);
     \vertex [left=0.3cm of b](c23);
     \vertex [above=0.23cm of c23](v23)[dot]{};
    \vertex [above=.5cm of c](j1){$1$};
    \vertex [right=.9cm of j1](j2){$\cdots n{-}3~~~~~$};
         \vertex [below=0.3cm of v23](l4){$n{-}2~~~~$};
    \vertex [right=0.6cm of j2](j3){${q}$};
   	 \diagram*{(a) -- [thick] (b),(b) -- [thick] (j1),(v23) -- [thick] (j2),(b)--[thick](j3)};
    \end{feynman}  
  \end{tikzpicture}  ,
  \end{align}
  where ``red box'' represents the heavy particles and $g'$ is the massive gluon.  The label $(n-2)$ on an internal line will be promoted to an external massless gluon after a series of cuts. 
  Using the expression for the HEFT BCJ numerator in \eqref{eq:recJ}, the residue on this cut is given by
  \begin{align} \label{eq:WpWp}
  	(1- \alpha' p_{{q}1\ldots n-3}^2) (-1)^{n-3}W'(1\ldots n{-}3 ,{q})\, p_{{q}}\mdot v \, .
  \end{align}
  By replacing $v$ with the polarisation vector $\veps^\perp_{n-1}$, which is the transverse mode of the massive gluon, we obtain the numerator for $(n-2)$ massless gluons and one massive gluon, $\npre(1\ldots n{-}3, {q},(n{-}1)_{g'})$. 
  
The BCJ numerator $\npre(1\ldots (n{-}3) {q},(n-1)_{g'})$ also factorises on the massless cut on the pole $1/ p^2_{12\ldots n{-}3}$. Consider the amplitude with $n{-}2$ massless gluons and one massive gluon (see more details in the Appendix \ref{app2}).  On this cut we have
\begin{align}
 \sum_{{\rm states}~ I} A(12\ldots n-3, \veps^*_I) \tr (F_I\cdot F_{{q}}) p_{{q}}\mdot\veps^{\perp}_{n-1}\, ,
\end{align}
where `$I$' denotes the internal state, and we have used the three point amplitude of two massless gluons and one massive gluon, which gives the trace factor. We now choose the polarisation vector $\veps_{{q}}$ to be orthogonal to the cut momentum, so the trace factor becomes 
\begin{align}
	2\alpha'\veps_I\mdot \veps_{{q}} p_I\mdot p_{{q}}=\veps_I\mdot \veps_{{q}}\, ,
\end{align}
where we used the massive on-shell condition of the external massive gluon. So on the cut and with the extra constraint on the polarisation vector $\veps_{{q}}$, the summation over the intermediate state gives a massless gluon amplitude
\begin{align} \label{eq:amm}
	&\sum_{{\rm states} I} A(1\ldots n{-}3, \veps^*_I) \veps_I\mdot \veps_{{q}} p_{{q}}\mdot\veps^{\perp}_{n-1}\nn\\
	&=A(1\ldots n-3, \veps_{{q}}) p_{{q}}\mdot\veps^{\perp}_{n-1}\, .
\end{align}
 Comparing \eqref{eq:amm} with \eqref{eq:WpWp} and taking $\veps_{n-2}=\veps_{{q}}$, we obtain an important relation between the function $W'(1\ldots (n{-}3) {q})$ and the BCJ numerator of massless gluons
\begin{align}\label{eq:factmassless}
&\npre(12\ldots n{-3}, n{-}2)=(-1)^{n-3} \, \times \\
	&\Big((1-\alpha' p_{1\ldots (n{-}3) {q}}^2)W'(1\ldots (n{-}3) {q})\Big)\Big|_{\rm cuts} \, , \nn
	\end{align}
 where the label `cuts' denotes all the above constraints \begin{align}\label{eq:cut}
     p_{1\ldots (n{-}3) {q}}^2-1/\alpha'=0,p_{1\ldots n-3}^2=p_{1\ldots n-3}\mdot \veps_{{q}}=0
 \end{align} 
With the BCJ numerator, the amplitude is then given by
\begin{align}
	A(1\ldots n{-}3, n{-}2)&=\sum_{\beta\in S_{n-4}} \mathrm{m}(1\ldots n{-}3, 1\beta)\, \npre(1\beta,n{-}2) \, .
\end{align}
Importantly, the amplitude $A(1\ldots n{-}3, n{-}2)$ can also be obtained by decoupling the heavy particles via a factorisation limit \cite{Brandhuber:2021bsf}, 
\begin{align}
    A(1\ldots n{-}3, n{-}2)=A(1\ldots n{-}3, v)|_{v\rightarrow \veps_{n-2}}^{p^2_{1\ldots n-3}=0}\, ,
\end{align}
where $A(1\ldots n{-}3, v)$ depends on $W'$-functions with $n{-}3$ or less gluons. In this way, we derive recursive relations of $W'$-functions.


\section{Combinatorial solution of the $W'$-function}

 We will now use the recursive relations discussed above and relabeling symmetry \eqref{eq:crossingW} to determine the $W'$-functions. We further impose manifest gauge invariance as well as factorization properties on massive  particle cuts. 

We first write down the general solution to the relabeling symmetry \eqref{eq:crossingW}. A particular solution of $W'(i_1\ldots i_{r-1} i_r)$ that obeys \eqref{eq:crossingW} was already constructed in \cite{Chen:2023ekh},   
\begin{align} \label{eq:W0}
W_0(i_1\ldots i_{r})\equiv \tr(F_{[i_1\ldots i_{r-1}]}F_{i_r}) \, ,
\end{align}
which is precisely the leading-$\alpha'$ correction term.  Two key properties of this particular solution are the left nested commutator of the indices except the last one and cyclic permutation invariance of the trace function. This observation leads to the general formal solution of \eqref{eq:crossingW}:  
\begin{align} \label{eq:operator}
	& \mathbb{O}_{{\rm cyc}([i_1\ldots i_{r-1}]i_r)} \circ f(i_1,i_2,\ldots i_{r})\equiv \nn\\
 &\sum_{\sigma \in [i_1\ldots i_{r-1}]i_r}\sum_{\rho_1 \rho_2\ldots \rho_r={\rm cyc}(\sigma)} f(\rho_1,\rho_2,\ldots,\rho_r)\, .
\end{align}

We then impose \eqref{eq:factmassless}. We proceed by defining partitions of ordered gluon indices $\{i_1, i_2,\ldots, i_{r-1}, i_r \}$ in $f(i_1, \ldots, i_r)$:
\begin{align}
\mathbf{P}_{i_1\ldots i_{r}}( x_1,x_2,\ldots,x_s) \, , \qquad  {\rm with} \qquad \sum_{i=1}^{s} x_i=r \, . 
\end{align}
For example, 
\begin{align}
   \mathbf{P}_{1234}( 2,1,1) &\equiv (12)|(3)|(4)\, , \label{eq:p1234} \\ 
  \mathbf{P}_{123456}( 2,1,2,1) &\equiv  (12)|(3)|(45)|(6) \, .  \label{eq:p123456}
\end{align}
To maintain manifest gauge invariance, each single-index subset, such as `$(3)$' and `$(4)$' in \eqref{eq:p1234}, is mapped to its corresponding strength tensor; in order to preserve the factorization behaviour on the massive cuts, each multi-index subset, such as `$(12)$' in \eqref{eq:p1234} and `$(45)$' in \eqref{eq:p123456}, is mapped to a lower-point $W'$-function. Power counting considerations then dictate that all single-index subsets between two multi-index subsets, for example,  
$   (i_L, \ldots, j_L) | (j), \ldots, (k) | (i_R, \ldots, j_R) \, , $
 must be mapped to $p_X \mdot F_j \mdot \ldots \mdot F_k \mdot p_Y$, or mapped to $p_X\mdot p_Y$ if there are no single-index subsets. To ensure the ordering of gluon indices, the sequence of dot products and traces involving field strengths must mirror the ordering of gluons. So, the labels $X$ and $Y$ can only correspond to adjacent indices in the lower-point $W'$-functions, i.e., $X=j_L, Y=i_R$. These rules establish a one-to-one map from each partition to a gauge-invariant function, e.g. for the partition given in \eqref{eq:p123456}, we have, 
\begin{align}
(12)|(3)|(45)|(6)
 = W'(12) p_2\mdot F_3\mdot p_4 W'(45) p_5 \mdot F_6\mdot p_1 \, . 
\end{align}
The single set partition and the maximal set partition are special, and are directly fixed by power counting 
\begin{align}
	(i_1i_2\ldots i_{r})&  = \tr(F_{i_1i_2\ldots i_{r}})\, ,\nn\\
	(i_1)|(i_2)|\ldots |(i_{r})& = 0 \,  .
\end{align}  
We are thus led to a simple and well-structured solution for the massless factorization behavior outlined in equation \eqref{eq:factmassless}. To incorporate the relabelling symmetry \eqref{eq:crossingW}, we sum over all partitions using the operator $\mathbb{O}$ defined in \eqref{eq:operator}. In conlcusion, we have, 
\begin{align}\label{eq:WResum}
    &W'(i_1\ldots i_{r-1} i_r)=\frac{\alpha'}{1-\alpha' p_{i_1\ldots i_{r}}^2}\Bigg[W_0(i_1\ldots i_{r-1} i_r)+\nn\\
    & \mathbb{O}_{{\rm cyc}([i_1\ldots i_{r-1}]i_r)} \circ \Big(\sum_{s=1}^{r-1}\sum_{ x_1+\cdots+ x_s=r}^{x_i\in \mathbb{Z}^+} {\mathbf{P}_{i_1\ldots i_{r}}( x_1,\ldots,  x_s)\over s}\Big)\Bigg],
\end{align}
where each partition is weighted by its corresponding overcounting number \footnote{Due to the cyclic summation in the operator $\mathbb{O}$, if ${\rm cyc}( x_1,\ldots,  x_s)= {\rm cyc}( x'_1,\ldots,  x'_s)$, the two  partitions  generate identical terms under the action of $\mathbb{O}_{{\rm cyc}([i_1\ldots i_{r-1}]i_r)}$. Therefore, for each partition element with $s$ subsets, the $\mathbb{O}$ operator introduces an overcounting factor equal to $s$.}, and we have identified $\mathbf{P}_{i_1\ldots i_{r}}( x_1,\ldots,  x_s)$ with lower-point $W'$-functions according to the rules we just discussed, 
\begin{align} \label{eq:defP}
	&\mathbf{P}_{i_1\ldots i_{r}}( x_1,\ldots,  x_s)=\nn\\
 &\bigg(\prod_{k=1}^{t-1} W'(i_{{\rm a}(k)}\ldots i_{b(k)})  p_{i_{b(k)}}\mdot F_{i_{b(k)+1}\ldots i_{a(k)-1}}\mdot p_{i_{a(k+1)}}\bigg)\nn\\
	&\times W'(i_{{\rm a}(t)}\ldots i_{b(t)})p_{i_{b(t)}}\mdot F_{i_{b(t)+1}\ldots r 1\cdots  i_{a(1)-1}}\mdot p_{i_{a(1)}}\, ,
\end{align}
where $t$ is the number of sets with multi gluon indices.   We have checked this general solution up to the eight-point HEFT amplitude (six gluons). Below are some simple examples of $W'$-functions (more examples can be found in Appendix \ref{app3}), 
 \begin{align}
 &W'(12)={\alpha'\over 1-\alpha'  p_{12}^2}  W_0(12) \, ,  \\
 	&W'(123)={\alpha'\over 1-\alpha'  p_{123}^2}  \Big(W_0(123)-2 W'(23) p_2\mdot F_1\mdot p_3\nn\\
 	&-2 W'(12) p_1\mdot F_3\mdot p_2+2 W'(13) p_1\mdot F_2\mdot p_3\Big)\, .
 \end{align}
with $W_0$ given in (\ref{eq:W0}). Note that the poles are the physical massive propagators in the $\rm DF^2+YM$ theory. In the $W'$-function, the number of terms in terms of $W_0$ and strength tensor products.
  \begin{center}
\begin{tabular}{||c||c| c| c| c| c| c|} 
 \hline
 $\textrm{gluons}$  & 2 & 3 & 4 & 5 & 6 & 7 
 \\ [0.5ex] 
 \hline
 $\#$ terms in $W'$  \, & \,1\,  & \,4\,  & \,45\, & \,921\, & \,30485\, & \,1539170\, 
 \, \\ 
 \hline
\end{tabular}
\end{center}

\section{Local BCJ numerator for the massless gluon amplitude}
As shown in \eqref{eq:factmassless}, the function $W'(1\ldots (n{-}3){q})$ automatically generates the BCJ numerators for the massless gluons. However, the numerators given in \eqref{eq:factmassless} in general contain spurious poles, due to the auxiliary momentum $p_{{q}}$. We denote this BCJ numerator as $\npre_{\rm NL}(12\ldots n-3, n{-}2)$, where `NL' stands for `non-local'. Expanding in terms of the $\alpha'$, we have
\begin{align} \label{eq:expansion}
	\npre_{\rm NL}(12\ldots n{-}3,n{-}2)&=\npre^{\rm YM}_{\rm NL}(12\ldots n{-}3, n{-}2)\\
	&+\sum_{\rm i=1}^{\infty}\alpha'\npre^{({\rm i})}_{\rm NL}(12\ldots n{-}3, n{-}2) \, . \nn
\end{align} 
where the expanded terms are non-local BCJ numerator for each order of $\alpha'$. 

We will now exploit the fact that result is independent of the auxiliary momentum $p_{{q}}$. To do so, we first impose the massive on-shell condition and massless cut condition in \eqref{eq:cut}, which gives, 
\begin{align}\label{eq:tachyonOnshell}
	 p_{1}\mdot p_{{q}}= 1/(2\alpha')- p_{2\ldots n-3}\mdot p_{{q}} \, ,
\end{align}
allowing us to remove $p_{1}\mdot p_{{q}}$, leaving only independent kinematic variables. 
The spurious poles in $\npre_{\rm NL}$ originate from the massive poles, and have the form
\begin{align}
    {1\over (\sum p_X\mdot p_{Y})+p_{Z}\mdot p_{{q}}}\, .
\end{align}
Since $p_{{q}}$ is auxiliary, the spurious pole is conveniently removed  by setting 
  \begin{align}\label{eq:localise}
      p_i\mdot p_{{q}}=z\rightarrow \infty ~{\rm for}~ i>1 \, . 
  \end{align}
Importantly, this choice preserves the relabeling symmetry of the BCJ numerator for $\{2,3,\cdots, n{-}3\}$, but spoils its gauge invariance. Applying this procedure to the $\alpha'$-expansion \eqref{eq:expansion}, we have
  \begin{align}
	\npre^{\rm (i)}(12\ldots n{-}3, n{-}2)=\npre^{\rm (i)}_{\rm NL}(12\ldots n{-}3, n{-}2)|^{p_i\mdot p_{{q}}=z}_{  z\rightarrow \infty},
\end{align} 
where the superscript index ${\rm i}={\rm YM}, 1,2,\cdots$. Now, the BCJ numerator does not contain any denominators and is purely local. Below are a few examples of such local BCJ numerators. At three points,  the local BCJ numerators from the $W'$-function are
\begin{align}\label{ym3pt}
	\npre^{\rm YM}(123)&=-\veps _1\mdot F_2\mdot \veps _3+\veps _1\mdot \veps _3 \veps _2\mdot p_1\, , \\
	\npre^{(1)}(123)&=-2 \alpha'  \veps _1\mdot p_2 \veps _3\mdot p_2 \veps _2\mdot p_1\, ,
\end{align}
with higher-order corrections vanishing. At four points, the local YM BCJ numerator is, 
\begin{align}\label{ym4pt}
	&\npre^{\rm YM}(1234)=\frac{1}{2} \veps _1\mdot \veps _4 \veps _2\mdot p_1 p_{12}\mdot \veps _3-\frac{1}{2} \veps _1\mdot \veps _4 p_1\mdot F_2\mdot \veps _3\nn\\
 &-\veps _2\mdot p_1 \veps _1\mdot F_3\mdot \veps _4+\frac{1}{2} \veps _1\mdot \veps _4 \veps _3\mdot p_1 \veps _2\mdot p_1-p_{12}\mdot \veps _3 \veps _1\mdot F_2\mdot \veps _4\nn\\
 &+\veps _1\mdot F_2\mdot F_3\mdot \veps _4 \, .
\end{align}
At four point this surprisingly matches the numerators in \cite{Du:2017kpo,Edison:2020ehu}, but at higher points the numerators differ by generalized gauge transformations, which only leave the amplitude invariant.
%
More examples are included in Appendix \ref{app4} and {\href{https://github.com/AmplitudeGravity/kinematicHopfAlgebra}{\it \blue KiHA5.0}}~\cite{ChenGitHub}.

\section{Kinematic algebra for local BCJ numerators in Yang-Mills theory}

We will show that for pure YM theory, the local BCJ numerators can also be obtained directly from the kinematic Hopf algebra, with a corresponding evaluation map
\begin{align} \label{eq:Nhat}
   \npre^{\rm YM}(12\ldots  n{-}2)=\langle T_{(1)}\star T_{(2)}\cdots \star T_{(n{-}3)}\rangle_{\rm YM} \, .
\end{align}
In this construction, the gluon $(n{-}2)$ will enter the mapping rule differently from the others, and does not have an associated generator.
The evaluation map can be deduced from the  $W'$-function when it is truncated to leading orders in the $\alpha'$ expansion. When expanding in $\alpha'$, the on-shell condition \eqref{eq:factmassless} for the massive gluon can reduce the $\alpha'$ order by one power. Since we are interested in pure YM, this implies we can focus purely on terms up to order linear in $\alpha'$, which are single trace terms of the form,
\begin{align}
    \alpha' W_0(1\ldots (n{-}3) {q})+ \sum \alpha'  p_{X}\mdot F_{\tau_b}\mdot p_{Y} W'(i_1\tau_ai_r)\,  .
\end{align}
On the massive gluon on-shell condition \eqref{eq:tachyonOnshell}, the first term contributes to leading order as 
\begin{align}
    \veps_1\mdot F_{2\ldots n-3}\mdot \veps_{{q}}\, .
\end{align}
The second term contributes to the leading order only when $i_1\tau_2i_r$ contains gluon indices $1$ and ${q}$, since other terms are independent under the on-shell condition \eqref{eq:tachyonOnshell}. Then, according to the $W'$-function relations  in eq.~\eqref{eq:crossingW}, the second term can be written as 
\begin{align}\label{eq:FWnum}
    \sum \alpha'  p_{X}\mdot F_{\tau_b}\mdot p_{{q}Y} W'(1\tau_a {q})\, .
\end{align}
 The propagator in the $W'$-function is also simplified for the leading order contribution in the limit $p_i\mdot p_{{q}}=z\rightarrow \infty$,
\begin{align}\label{eq:FWden}
    {\alpha'\over 1-\alpha'p_{{q}1\tau_a}^2}\rightarrow {1\over 2p_{{q}}\mdot p_{\tau_b}}\, ,
\end{align}
where we used the on-shell condition $p_{{q}1\tau_a\tau_b}^2=1/\alpha'$.

In the  limit $z\rightarrow \infty$,   the parameter $z$ in \eqref{eq:FWnum} and \eqref{eq:FWden} cancels, which ensures that the BCJ numerator retains its local form. We can now identify the specific terms within the $W'$-function that contribute to local BCJ numerators, leading to the evaluation map: 
\begin{align}\label{eq:T}
&\langle T_{(1\tau_1),(\tau_2),\ldots,(\tau_r)}\rangle_{\rm YM}:=\begin{tikzpicture}[baseline={([yshift=-0.8ex]current bounding box.center)}]\tikzstyle{every node}=[font=\small]   
   \begin{feynman}
    \vertex (a)[myblob]{};
     \vertex[right=0.8cm of a] (a2)[myblob]{};
      \vertex[right=0.8cm of a2] (a3)[myblob]{};
       \vertex[right=0.8cm of a3] (a4)[myblob]{};
       \vertex[above=0.8cm of a] (b1){$1\tau_1~$};
        \vertex[above=0.8cm of a2] (b2){$\tau_2$};
        \vertex[above=0.8cm of a3] (b3){$\cdots$};
         \vertex[above=0.8cm of a4] (b4){$\tau_r$};
       \vertex [above=0.8cm of a](j1){$ $};
    \vertex [left=0.35cm of j1](j2){$ $};
    \vertex [right=0.55cm of j2](j3){$ $};
    \vertex [right=0.4cm of j3](j4){$ $};
    \vertex [right=0.6cm of j4](j5){$ $};
      \vertex [right=0.2cm of j5](j6){$ $};
    \vertex [right=0.6cm of j6](j7){$ $};
     \vertex [right=0.0cm of j7](j8){$ $};
    \vertex [right=0.8cm of j8](j9){$ $};
  	 \diagram*{(a)--[very thick](a2),(a2)--[very thick](a3), (a3)--[very thick](a4),(a) -- [thick] (j2),(a)--[thick](j3),(a2) -- [thick] (j4),(a2)--[thick](j5),(a4) -- [thick] (j8),(a4)--[thick](j9)};
    \end{feynman}  
  \end{tikzpicture}=\nn\\
& {\sc G_{1\tau_1}}  {\sc G_{\tau_2}(p_{ \Theta(\tau_{2})})\over {n{-}3{-}| {\tau_1}}|}\, \cdots {\sc G_{\tau_r}(p_{\Theta(\tau_{r})})\over {n{-}3{-}|{\tau_1\ldots \tau_{r-1}}|}} \, ,  \nn\\
&\langle T_{(j\tau_1),(\tau_2),\ldots,(\tau_r)}\rangle_{\rm YM}:=0, ~~~\text{if}~~~ j>1\, , 
\end{align}
where each denominator contains a symmetry factor, and
\begin{align}
	&\sc G_{1\tau_1}=\veps_1\mdot F_{\tau_1}\mdot \veps_{n-2}\, , \nn\\
	&\sc G_{\tau_ij}(p_{\Theta(\tau_{i}j)})=p_{\Theta(\tau_ij)}\mdot F_{\tau_i}\mdot \veps_{j}\, .
\end{align}
 The set $ \Theta(\tau_i)$ consists of all indices to the left of $\tau_i$ and smaller than the first  index in $\tau_i$, as in \eqref{eq:recJ}. 
This reproduces for example equations (\ref{ym3pt}) and (\ref{ym4pt}), demonstrating that indeed Hopf algebras can apply directly to YM, without relying on factorisation limits of HEFT numerators. We note the number of terms in the local BCJ numerators are twice of that in the non-local BCJ numerator \cite{Brandhuber:2021bsf} (i.e. Fubini numbers \cite{10.2307/2312725}), the same as what was obtained in \cite{Cheung:2021zvb} using different methods. This approach can be further extended to higher $\alpha'$ corrections of the local BCJ numerators by generalising the above analysis. 

\section{Conclusion and outlook}
This letter investigates the construction of the BCJ numerators for $\rm DF^2+YM$ theory, whose $\alpha'$ expansion generates higher derivative corrections to YM theory. The approach is based on the kinematic Hopf algebra, where a fusion product of generators ensures the color-kinematic duality holds, with the full $\alpha'$ dependence contained in the mapping rule from abstract generators to kinematic functions. We also derive the evaluation map that gives a local and crossing symmetric expression for YM BCJ numerators. 
In this construction, it is crucial to note that the quasi-shuffle product encodes factorization properties of massive poles at this numerator level. This is indeed consistent with the quasi-shuffle Hopf algebra. Let us take the simplest example, $T_{(1)}\star T_{(2)}=\overbrace{T_{(1),(2)}+T_{(2),(1)}}^{\text{shuffle}}-\overbrace{T_{(12)}}^{\text{stuffing}}\, ,$    
where the shuffle part precisely gives the factorization behaviour. The factorizing pieces alone are not sufficient for the numerator to obey colour-kinematic duality, and indeed the ``stuffing term'' gives the necessary correction, explaining the need for a quasi-shuffle, instead of a simple shuffle product.  

The appearance of Hopf algebras in this context further suggests this structure universally underpins the color-kinematic duality. To test this fascinating possibility, by using transmutation operators  \cite{Cheung:2017ems} or dimensional reduction \cite{Chiodaroli:2015rdg}, one could also explore the appearance of Hopf algebras in other theories of scalars with higher derivatives, and connect to different approaches such as \cite{Carrasco:2019yyn,Low:2020ubn,Carrasco:2021ptp,Pavao:2022kog}.

Recently it was shown that  $\rm DF^2+YM$  theory is not the only theory containing potential higher derivative corrections to YM, but instead there exists an infinite family of such theories \cite{Carrasco:2022sck}. Interestingly, we find that all such other theories are order $\mathcal{O}(m^3)$ or higher in the HEFT limit, corresponding to hyper-classical observables, whereas  $\rm DF^2+YM$ theory is order $\mathcal{O}(m)$. It would be interesting to explore whether these other higher-derivative theories can also be described by a kinematic Hopf algebra. 


The BCJ numerators presented in this Letter serve as a direct means to construct gravitational amplitudes extending beyond Einstein gravity \cite{Brandhuber:2019qpg,AccettulliHuber:2020oou,AccettulliHuber:2020dal,Emond:2019crr,Sennett:2019bpc}. By incorporating them into the classical HEFT expansion  graphs \cite{Brandhuber:2021eyq,Brandhuber:2023hhy,Herderschee:2023fxh} of binary black hole scattering, our approach facilitates the calculation of classical observables such as bending angles or waveforms within the binary black hole system, enabling the study of potential physical effects beyond pure Einstein gravity.

\section{Acknowledgements}
We would like to thank Andreas Brandhuber and Gabriele Travaglini for stimulating discussions.  G.C. has received funding from the European Union's Horizon 2020 research and innovation program under the Marie Sk\l{}odowska-Curie grant agreement No.~847523 ``INTERACTIONS''.  C.W. is supported by a Royal Society University Research Fellowship URF\textbackslash R\textbackslash 221015 and a STFC Consolidated Grant, ST\textbackslash T000686\textbackslash 1 ``Amplitudes, strings \& duality".

\bibliographystyle{apsrev4-1}
\bibliography{KinematicAlgebra}

\begin{thebibliography}{73}%
\makeatletter
\providecommand \@ifxundefined [1]{%
 \@ifx{#1\undefined}
}%
\providecommand \@ifnum [1]{%
 \ifnum #1\expandafter \@firstoftwo
 \else \expandafter \@secondoftwo
 \fi
}%
\providecommand \@ifx [1]{%
 \ifx #1\expandafter \@firstoftwo
 \else \expandafter \@secondoftwo
 \fi
}%
\providecommand \natexlab [1]{#1}%
\providecommand \enquote  [1]{``#1''}%
\providecommand \bibnamefont  [1]{#1}%
\providecommand \bibfnamefont [1]{#1}%
\providecommand \citenamefont [1]{#1}%
\providecommand \href@noop [0]{\@secondoftwo}%
\providecommand \href [0]{\begingroup \@sanitize@url \@href}%
\providecommand \@href[1]{\@@startlink{#1}\@@href}%
\providecommand \@@href[1]{\endgroup#1\@@endlink}%
\providecommand \@sanitize@url [0]{\catcode `\\12\catcode `\$12\catcode
  `\&12\catcode `\#12\catcode `\^12\catcode `\_12\catcode `\%12\relax}%
\providecommand \@@startlink[1]{}%
\providecommand \@@endlink[0]{}%
\providecommand \url  [0]{\begingroup\@sanitize@url \@url }%
\providecommand \@url [1]{\endgroup\@href {#1}{\urlprefix }}%
\providecommand \urlprefix  [0]{URL }%
\providecommand \Eprint [0]{\href }%
\providecommand \doibase [0]{http://dx.doi.org/}%
\providecommand \selectlanguage [0]{\@gobble}%
\providecommand \bibinfo  [0]{\@secondoftwo}%
\providecommand \bibfield  [0]{\@secondoftwo}%
\providecommand \translation [1]{[#1]}%
\providecommand \BibitemOpen [0]{}%
\providecommand \bibitemStop [0]{}%
\providecommand \bibitemNoStop [0]{.\EOS\space}%
\providecommand \EOS [0]{\spacefactor3000\relax}%
\providecommand \BibitemShut  [1]{\csname bibitem#1\endcsname}%
\let\auto@bib@innerbib\@empty
\bibitem [{\citenamefont {Bern}\ \emph {et~al.}(2008)\citenamefont {Bern},
  \citenamefont {Carrasco},\ and\ \citenamefont {Johansson}}]{Bern:2008qj}%
  \BibitemOpen
  \bibfield  {author} {\bibinfo {author} {\bibfnamefont {Z.}~\bibnamefont
  {Bern}}, \bibinfo {author} {\bibfnamefont {J.~J.~M.}\ \bibnamefont
  {Carrasco}}, \ and\ \bibinfo {author} {\bibfnamefont {H.}~\bibnamefont
  {Johansson}},\ }\href {\doibase 10.1103/PhysRevD.78.085011} {\bibfield
  {journal} {\bibinfo  {journal} {Phys. Rev.}\ }\textbf {\bibinfo {volume}
  {D78}},\ \bibinfo {pages} {085011} (\bibinfo {year} {2008})},\ \Eprint
  {http://arxiv.org/abs/0805.3993} {arXiv:0805.3993 [hep-ph]} \BibitemShut
  {NoStop}%
\bibitem [{\citenamefont {Bern}\ \emph {et~al.}(2010)\citenamefont {Bern},
  \citenamefont {Carrasco},\ and\ \citenamefont {Johansson}}]{Bern:2010ue}%
  \BibitemOpen
  \bibfield  {author} {\bibinfo {author} {\bibfnamefont {Z.}~\bibnamefont
  {Bern}}, \bibinfo {author} {\bibfnamefont {J.~J.~M.}\ \bibnamefont
  {Carrasco}}, \ and\ \bibinfo {author} {\bibfnamefont {H.}~\bibnamefont
  {Johansson}},\ }\href {\doibase 10.1103/PhysRevLett.105.061602} {\bibfield
  {journal} {\bibinfo  {journal} {Phys. Rev. Lett.}\ }\textbf {\bibinfo
  {volume} {105}},\ \bibinfo {pages} {061602} (\bibinfo {year} {2010})},\
  \Eprint {http://arxiv.org/abs/1004.0476} {arXiv:1004.0476 [hep-th]}
  \BibitemShut {NoStop}%
\bibitem [{\citenamefont {Monteiro}\ and\ \citenamefont
  {O'Connell}(2011)}]{Monteiro:2011pc}%
  \BibitemOpen
  \bibfield  {author} {\bibinfo {author} {\bibfnamefont {R.}~\bibnamefont
  {Monteiro}}\ and\ \bibinfo {author} {\bibfnamefont {D.}~\bibnamefont
  {O'Connell}},\ }\href {\doibase 10.1007/JHEP07(2011)007} {\bibfield
  {journal} {\bibinfo  {journal} {JHEP}\ }\textbf {\bibinfo {volume} {07}},\
  \bibinfo {pages} {007} (\bibinfo {year} {2011})},\ \Eprint
  {http://arxiv.org/abs/1105.2565} {arXiv:1105.2565 [hep-th]} \BibitemShut
  {NoStop}%
\bibitem [{\citenamefont {Bjerrum-Bohr}\ \emph {et~al.}(2012)\citenamefont
  {Bjerrum-Bohr}, \citenamefont {Damgaard}, \citenamefont {Monteiro},\ and\
  \citenamefont {O'Connell}}]{Bjerrum-Bohr:2012kaa}%
  \BibitemOpen
  \bibfield  {author} {\bibinfo {author} {\bibfnamefont {N.~E.~J.}\
  \bibnamefont {Bjerrum-Bohr}}, \bibinfo {author} {\bibfnamefont {P.~H.}\
  \bibnamefont {Damgaard}}, \bibinfo {author} {\bibfnamefont {R.}~\bibnamefont
  {Monteiro}}, \ and\ \bibinfo {author} {\bibfnamefont {D.}~\bibnamefont
  {O'Connell}},\ }\href {\doibase 10.1007/JHEP06(2012)061} {\bibfield
  {journal} {\bibinfo  {journal} {JHEP}\ }\textbf {\bibinfo {volume} {06}},\
  \bibinfo {pages} {061} (\bibinfo {year} {2012})},\ \Eprint
  {http://arxiv.org/abs/1203.0944} {arXiv:1203.0944 [hep-th]} \BibitemShut
  {NoStop}%
\bibitem [{\citenamefont {Cheung}\ and\ \citenamefont
  {Shen}(2017)}]{Cheung:2016prv}%
  \BibitemOpen
  \bibfield  {author} {\bibinfo {author} {\bibfnamefont {C.}~\bibnamefont
  {Cheung}}\ and\ \bibinfo {author} {\bibfnamefont {C.-H.}\ \bibnamefont
  {Shen}},\ }\href {\doibase 10.1103/PhysRevLett.118.121601} {\bibfield
  {journal} {\bibinfo  {journal} {Phys. Rev. Lett.}\ }\textbf {\bibinfo
  {volume} {118}},\ \bibinfo {pages} {121601} (\bibinfo {year} {2017})},\
  \Eprint {http://arxiv.org/abs/1612.00868} {arXiv:1612.00868 [hep-th]}
  \BibitemShut {NoStop}%
\bibitem [{\citenamefont {Chen}\ \emph {et~al.}(2019)\citenamefont {Chen},
  \citenamefont {Johansson}, \citenamefont {Teng},\ and\ \citenamefont
  {Wang}}]{Chen:2019ywi}%
  \BibitemOpen
  \bibfield  {author} {\bibinfo {author} {\bibfnamefont {G.}~\bibnamefont
  {Chen}}, \bibinfo {author} {\bibfnamefont {H.}~\bibnamefont {Johansson}},
  \bibinfo {author} {\bibfnamefont {F.}~\bibnamefont {Teng}}, \ and\ \bibinfo
  {author} {\bibfnamefont {T.}~\bibnamefont {Wang}},\ }\href {\doibase
  10.1007/JHEP11(2019)055} {\bibfield  {journal} {\bibinfo  {journal} {JHEP}\
  }\textbf {\bibinfo {volume} {11}},\ \bibinfo {pages} {055} (\bibinfo {year}
  {2019})},\ \Eprint {http://arxiv.org/abs/1906.10683} {arXiv:1906.10683
  [hep-th]} \BibitemShut {NoStop}%
\bibitem [{\citenamefont {Mizera}(2020)}]{Mizera:2019blq}%
  \BibitemOpen
  \bibfield  {author} {\bibinfo {author} {\bibfnamefont {S.}~\bibnamefont
  {Mizera}},\ }\href {\doibase 10.1103/PhysRevLett.124.141601} {\bibfield
  {journal} {\bibinfo  {journal} {Phys. Rev. Lett.}\ }\textbf {\bibinfo
  {volume} {124}},\ \bibinfo {pages} {141601} (\bibinfo {year} {2020})},\
  \Eprint {http://arxiv.org/abs/1912.03397} {arXiv:1912.03397 [hep-th]}
  \BibitemShut {NoStop}%
\bibitem [{\citenamefont {Chen}\ \emph {et~al.}(2021)\citenamefont {Chen},
  \citenamefont {Johansson}, \citenamefont {Teng},\ and\ \citenamefont
  {Wang}}]{Chen:2021chy}%
  \BibitemOpen
  \bibfield  {author} {\bibinfo {author} {\bibfnamefont {G.}~\bibnamefont
  {Chen}}, \bibinfo {author} {\bibfnamefont {H.}~\bibnamefont {Johansson}},
  \bibinfo {author} {\bibfnamefont {F.}~\bibnamefont {Teng}}, \ and\ \bibinfo
  {author} {\bibfnamefont {T.}~\bibnamefont {Wang}},\ }\href {\doibase
  10.1007/JHEP10(2021)042} {\bibfield  {journal} {\bibinfo  {journal} {JHEP}\
  }\textbf {\bibinfo {volume} {10}},\ \bibinfo {pages} {042} (\bibinfo {year}
  {2021})},\ \Eprint {http://arxiv.org/abs/2104.12726} {arXiv:2104.12726
  [hep-th]} \BibitemShut {NoStop}%
\bibitem [{\citenamefont {Ben-Shahar}\ and\ \citenamefont
  {Johansson}(2022)}]{Ben-Shahar:2021zww}%
  \BibitemOpen
  \bibfield  {author} {\bibinfo {author} {\bibfnamefont {M.}~\bibnamefont
  {Ben-Shahar}}\ and\ \bibinfo {author} {\bibfnamefont {H.}~\bibnamefont
  {Johansson}},\ }\href {\doibase 10.1007/JHEP08(2022)035} {\bibfield
  {journal} {\bibinfo  {journal} {JHEP}\ }\textbf {\bibinfo {volume} {08}},\
  \bibinfo {pages} {035} (\bibinfo {year} {2022})},\ \Eprint
  {http://arxiv.org/abs/2112.11452} {arXiv:2112.11452 [hep-th]} \BibitemShut
  {NoStop}%
\bibitem [{\citenamefont {Monteiro}\ \emph {et~al.}(2023)\citenamefont
  {Monteiro}, \citenamefont {Stark-Much\~ao},\ and\ \citenamefont
  {Wikeley}}]{Monteiro:2022nqt}%
  \BibitemOpen
  \bibfield  {author} {\bibinfo {author} {\bibfnamefont {R.}~\bibnamefont
  {Monteiro}}, \bibinfo {author} {\bibfnamefont {R.}~\bibnamefont
  {Stark-Much\~ao}}, \ and\ \bibinfo {author} {\bibfnamefont {S.}~\bibnamefont
  {Wikeley}},\ }\href {\doibase 10.1007/JHEP09(2023)030} {\bibfield  {journal}
  {\bibinfo  {journal} {JHEP}\ }\textbf {\bibinfo {volume} {09}},\ \bibinfo
  {pages} {030} (\bibinfo {year} {2023})},\ \Eprint
  {http://arxiv.org/abs/2211.12407} {arXiv:2211.12407 [hep-th]} \BibitemShut
  {NoStop}%
\bibitem [{\citenamefont {Monteiro}(2023)}]{Monteiro:2022lwm}%
  \BibitemOpen
  \bibfield  {author} {\bibinfo {author} {\bibfnamefont {R.}~\bibnamefont
  {Monteiro}},\ }\href {\doibase 10.1007/JHEP01(2023)092} {\bibfield  {journal}
  {\bibinfo  {journal} {JHEP}\ }\textbf {\bibinfo {volume} {01}},\ \bibinfo
  {pages} {092} (\bibinfo {year} {2023})},\ \Eprint
  {http://arxiv.org/abs/2208.11179} {arXiv:2208.11179 [hep-th]} \BibitemShut
  {NoStop}%
\bibitem [{\citenamefont {Lipstein}\ and\ \citenamefont
  {Nagy}(2023)}]{Lipstein:2023pih}%
  \BibitemOpen
  \bibfield  {author} {\bibinfo {author} {\bibfnamefont {A.}~\bibnamefont
  {Lipstein}}\ and\ \bibinfo {author} {\bibfnamefont {S.}~\bibnamefont
  {Nagy}},\ }\href {\doibase 10.1103/PhysRevLett.131.081501} {\bibfield
  {journal} {\bibinfo  {journal} {Phys. Rev. Lett.}\ }\textbf {\bibinfo
  {volume} {131}},\ \bibinfo {pages} {081501} (\bibinfo {year} {2023})},\
  \Eprint {http://arxiv.org/abs/2304.07141} {arXiv:2304.07141 [hep-th]}
  \BibitemShut {NoStop}%
\bibitem [{\citenamefont {Ferrero}\ and\ \citenamefont
  {Francia}(2021)}]{Ferrero:2020vww}%
  \BibitemOpen
  \bibfield  {author} {\bibinfo {author} {\bibfnamefont {P.}~\bibnamefont
  {Ferrero}}\ and\ \bibinfo {author} {\bibfnamefont {D.}~\bibnamefont
  {Francia}},\ }\href {\doibase 10.1007/JHEP02(2021)213} {\bibfield  {journal}
  {\bibinfo  {journal} {JHEP}\ }\textbf {\bibinfo {volume} {02}},\ \bibinfo
  {pages} {213} (\bibinfo {year} {2021})},\ \Eprint
  {http://arxiv.org/abs/2012.00713} {arXiv:2012.00713 [hep-th]} \BibitemShut
  {NoStop}%
\bibitem [{\citenamefont {Ben-Shahar}\ \emph {et~al.}(2023)\citenamefont
  {Ben-Shahar}, \citenamefont {Garozzo},\ and\ \citenamefont
  {Johansson}}]{Ben-Shahar:2022ixa}%
  \BibitemOpen
  \bibfield  {author} {\bibinfo {author} {\bibfnamefont {M.}~\bibnamefont
  {Ben-Shahar}}, \bibinfo {author} {\bibfnamefont {L.}~\bibnamefont {Garozzo}},
  \ and\ \bibinfo {author} {\bibfnamefont {H.}~\bibnamefont {Johansson}},\
  }\href {\doibase 10.1007/JHEP08(2023)222} {\bibfield  {journal} {\bibinfo
  {journal} {JHEP}\ }\textbf {\bibinfo {volume} {08}},\ \bibinfo {pages} {222}
  (\bibinfo {year} {2023})},\ \Eprint {http://arxiv.org/abs/2301.00233}
  {arXiv:2301.00233 [hep-th]} \BibitemShut {NoStop}%
\bibitem [{\citenamefont {Mafra}\ and\ \citenamefont
  {Schlotterer}(2023)}]{Mafra:2022wml}%
  \BibitemOpen
  \bibfield  {author} {\bibinfo {author} {\bibfnamefont {C.~R.}\ \bibnamefont
  {Mafra}}\ and\ \bibinfo {author} {\bibfnamefont {O.}~\bibnamefont
  {Schlotterer}},\ }\href {\doibase 10.1016/j.physrep.2023.04.001} {\bibfield
  {journal} {\bibinfo  {journal} {Phys. Rept.}\ }\textbf {\bibinfo {volume}
  {1020}},\ \bibinfo {pages} {1} (\bibinfo {year} {2023})},\ \Eprint
  {http://arxiv.org/abs/2210.14241} {arXiv:2210.14241 [hep-th]} \BibitemShut
  {NoStop}%
\bibitem [{\citenamefont {Fu}\ \emph {et~al.}(2018)\citenamefont {Fu},
  \citenamefont {Vanhove},\ and\ \citenamefont {Wang}}]{Fu:2018hpu}%
  \BibitemOpen
  \bibfield  {author} {\bibinfo {author} {\bibfnamefont {C.-H.}\ \bibnamefont
  {Fu}}, \bibinfo {author} {\bibfnamefont {P.}~\bibnamefont {Vanhove}}, \ and\
  \bibinfo {author} {\bibfnamefont {Y.}~\bibnamefont {Wang}},\ }\href {\doibase
  10.1007/JHEP09(2018)141} {\bibfield  {journal} {\bibinfo  {journal} {JHEP}\
  }\textbf {\bibinfo {volume} {09}},\ \bibinfo {pages} {141} (\bibinfo {year}
  {2018})},\ \Eprint {http://arxiv.org/abs/1806.09584} {arXiv:1806.09584
  [hep-th]} \BibitemShut {NoStop}%
\bibitem [{\citenamefont {Ben-Shahar}\ and\ \citenamefont
  {Guillen}(2021)}]{Ben-Shahar:2021doh}%
  \BibitemOpen
  \bibfield  {author} {\bibinfo {author} {\bibfnamefont {M.}~\bibnamefont
  {Ben-Shahar}}\ and\ \bibinfo {author} {\bibfnamefont {M.}~\bibnamefont
  {Guillen}},\ }\href {\doibase 10.1007/JHEP12(2021)014} {\bibfield  {journal}
  {\bibinfo  {journal} {JHEP}\ }\textbf {\bibinfo {volume} {12}},\ \bibinfo
  {pages} {014} (\bibinfo {year} {2021})},\ \Eprint
  {http://arxiv.org/abs/2108.11708} {arXiv:2108.11708 [hep-th]} \BibitemShut
  {NoStop}%
\bibitem [{\citenamefont {Borsten}\ \emph {et~al.}(2021)\citenamefont
  {Borsten}, \citenamefont {Jur\v{c}o}, \citenamefont {Kim}, \citenamefont
  {Macrelli}, \citenamefont {Saemann},\ and\ \citenamefont
  {Wolf}}]{Borsten:2020zgj}%
  \BibitemOpen
  \bibfield  {author} {\bibinfo {author} {\bibfnamefont {L.}~\bibnamefont
  {Borsten}}, \bibinfo {author} {\bibfnamefont {B.}~\bibnamefont {Jur\v{c}o}},
  \bibinfo {author} {\bibfnamefont {H.}~\bibnamefont {Kim}}, \bibinfo {author}
  {\bibfnamefont {T.}~\bibnamefont {Macrelli}}, \bibinfo {author}
  {\bibfnamefont {C.}~\bibnamefont {Saemann}}, \ and\ \bibinfo {author}
  {\bibfnamefont {M.}~\bibnamefont {Wolf}},\ }\href {\doibase
  10.1103/PhysRevLett.126.191601} {\bibfield  {journal} {\bibinfo  {journal}
  {Phys. Rev. Lett.}\ }\textbf {\bibinfo {volume} {126}},\ \bibinfo {pages}
  {191601} (\bibinfo {year} {2021})},\ \Eprint
  {http://arxiv.org/abs/2007.13803} {arXiv:2007.13803 [hep-th]} \BibitemShut
  {NoStop}%
\bibitem [{\citenamefont {Borsten}\ and\ \citenamefont
  {Nagy}(2020)}]{Borsten:2020xbt}%
  \BibitemOpen
  \bibfield  {author} {\bibinfo {author} {\bibfnamefont {L.}~\bibnamefont
  {Borsten}}\ and\ \bibinfo {author} {\bibfnamefont {S.}~\bibnamefont {Nagy}},\
  }\href {\doibase 10.1007/JHEP07(2020)093} {\bibfield  {journal} {\bibinfo
  {journal} {JHEP}\ }\textbf {\bibinfo {volume} {07}},\ \bibinfo {pages} {093}
  (\bibinfo {year} {2020})},\ \Eprint {http://arxiv.org/abs/2004.14945}
  {arXiv:2004.14945 [hep-th]} \BibitemShut {NoStop}%
\bibitem [{\citenamefont {Bonezzi}\ \emph {et~al.}(2023)\citenamefont
  {Bonezzi}, \citenamefont {Diaz-Jaramillo},\ and\ \citenamefont
  {Nagy}}]{Bonezzi:2023pox}%
  \BibitemOpen
  \bibfield  {author} {\bibinfo {author} {\bibfnamefont {R.}~\bibnamefont
  {Bonezzi}}, \bibinfo {author} {\bibfnamefont {F.}~\bibnamefont
  {Diaz-Jaramillo}}, \ and\ \bibinfo {author} {\bibfnamefont {S.}~\bibnamefont
  {Nagy}},\ }\href {\doibase 10.1103/PhysRevD.108.065007} {\bibfield  {journal}
  {\bibinfo  {journal} {Phys. Rev. D}\ }\textbf {\bibinfo {volume} {108}},\
  \bibinfo {pages} {065007} (\bibinfo {year} {2023})},\ \Eprint
  {http://arxiv.org/abs/2306.08558} {arXiv:2306.08558 [hep-th]} \BibitemShut
  {NoStop}%
\bibitem [{\citenamefont {Borsten}\ \emph {et~al.}(2023)\citenamefont
  {Borsten}, \citenamefont {Jurco}, \citenamefont {Kim}, \citenamefont
  {Macrelli}, \citenamefont {Saemann},\ and\ \citenamefont
  {Wolf}}]{Borsten:2022vtg}%
  \BibitemOpen
  \bibfield  {author} {\bibinfo {author} {\bibfnamefont {L.}~\bibnamefont
  {Borsten}}, \bibinfo {author} {\bibfnamefont {B.}~\bibnamefont {Jurco}},
  \bibinfo {author} {\bibfnamefont {H.}~\bibnamefont {Kim}}, \bibinfo {author}
  {\bibfnamefont {T.}~\bibnamefont {Macrelli}}, \bibinfo {author}
  {\bibfnamefont {C.}~\bibnamefont {Saemann}}, \ and\ \bibinfo {author}
  {\bibfnamefont {M.}~\bibnamefont {Wolf}},\ }\href {\doibase
  10.1103/PhysRevLett.131.041603} {\bibfield  {journal} {\bibinfo  {journal}
  {Phys. Rev. Lett.}\ }\textbf {\bibinfo {volume} {131}},\ \bibinfo {pages}
  {041603} (\bibinfo {year} {2023})},\ \Eprint
  {http://arxiv.org/abs/2211.13261} {arXiv:2211.13261 [hep-th]} \BibitemShut
  {NoStop}%
\bibitem [{\citenamefont {Armstrong-Williams}\ \emph
  {et~al.}(2024)\citenamefont {Armstrong-Williams}, \citenamefont {Nagy},
  \citenamefont {White},\ and\ \citenamefont
  {Wikeley}}]{Armstrong-Williams:2024icu}%
  \BibitemOpen
  \bibfield  {author} {\bibinfo {author} {\bibfnamefont {K.}~\bibnamefont
  {Armstrong-Williams}}, \bibinfo {author} {\bibfnamefont {S.}~\bibnamefont
  {Nagy}}, \bibinfo {author} {\bibfnamefont {C.~D.}\ \bibnamefont {White}}, \
  and\ \bibinfo {author} {\bibfnamefont {S.}~\bibnamefont {Wikeley}},\
  }\href@noop {} {\  (\bibinfo {year} {2024})},\ \Eprint
  {http://arxiv.org/abs/2401.10750} {arXiv:2401.10750 [hep-th]} \BibitemShut
  {NoStop}%
\bibitem [{\citenamefont {Brown}\ \emph {et~al.}(2024)\citenamefont {Brown},
  \citenamefont {Gowdy},\ and\ \citenamefont {Spence}}]{Brown:2023zxm}%
  \BibitemOpen
  \bibfield  {author} {\bibinfo {author} {\bibfnamefont {G.~R.}\ \bibnamefont
  {Brown}}, \bibinfo {author} {\bibfnamefont {J.}~\bibnamefont {Gowdy}}, \ and\
  \bibinfo {author} {\bibfnamefont {B.}~\bibnamefont {Spence}},\ }\href
  {\doibase 10.1103/PhysRevD.109.026009} {\bibfield  {journal} {\bibinfo
  {journal} {Phys. Rev. D}\ }\textbf {\bibinfo {volume} {109}},\ \bibinfo
  {pages} {026009} (\bibinfo {year} {2024})},\ \Eprint
  {http://arxiv.org/abs/2307.11063} {arXiv:2307.11063 [hep-th]} \BibitemShut
  {NoStop}%
\bibitem [{\citenamefont {Fu}\ and\ \citenamefont {Wang}(2022)}]{Fu:2022esi}%
  \BibitemOpen
  \bibfield  {author} {\bibinfo {author} {\bibfnamefont {C.-H.}\ \bibnamefont
  {Fu}}\ and\ \bibinfo {author} {\bibfnamefont {Y.}~\bibnamefont {Wang}},\
  }\href {\doibase 10.1007/978-981-19-4751-3_23} {\bibfield  {journal}
  {\bibinfo  {journal} {Springer Proc. Math. Stat.}\ }\textbf {\bibinfo
  {volume} {396}},\ \bibinfo {pages} {287} (\bibinfo {year} {2022})},\ \Eprint
  {http://arxiv.org/abs/2310.19724} {arXiv:2310.19724 [hep-th]} \BibitemShut
  {NoStop}%
\bibitem [{\citenamefont {Hoffman}(2000)}]{hoffman2000quasi}%
  \BibitemOpen
  \bibfield  {author} {\bibinfo {author} {\bibfnamefont {M.~E.}\ \bibnamefont
  {Hoffman}},\ }\href@noop {} {\bibfield  {journal} {\bibinfo  {journal}
  {Journal of Algebraic Combinatorics}\ }\textbf {\bibinfo {volume} {11}},\
  \bibinfo {pages} {49} (\bibinfo {year} {2000})}\BibitemShut {NoStop}%
\bibitem [{\citenamefont {Blumlein}(2004)}]{Blumlein:2003gb}%
  \BibitemOpen
  \bibfield  {author} {\bibinfo {author} {\bibfnamefont {J.}~\bibnamefont
  {Blumlein}},\ }\href {\doibase 10.1016/j.cpc.2003.12.004} {\bibfield
  {journal} {\bibinfo  {journal} {Comput. Phys. Commun.}\ }\textbf {\bibinfo
  {volume} {159}},\ \bibinfo {pages} {19} (\bibinfo {year} {2004})},\ \Eprint
  {http://arxiv.org/abs/hep-ph/0311046} {arXiv:hep-ph/0311046} \BibitemShut
  {NoStop}%
\bibitem [{\citenamefont {Aguiar}\ and\ \citenamefont
  {Mahajan}(2010)}]{aguiar2010monoidal}%
  \BibitemOpen
  \bibfield  {author} {\bibinfo {author} {\bibfnamefont {M.}~\bibnamefont
  {Aguiar}}\ and\ \bibinfo {author} {\bibfnamefont {S.~A.}\ \bibnamefont
  {Mahajan}},\ }\href@noop {} {\emph {\bibinfo {title} {Monoidal functors,
  species and Hopf algebras}}},\ Vol.~\bibinfo {volume} {29}\ (\bibinfo
  {publisher} {American Mathematical Society Providence, RI},\ \bibinfo {year}
  {2010})\BibitemShut {NoStop}%
\bibitem [{\citenamefont {Hoffman}\ and\ \citenamefont
  {Ihara}(2017)}]{hoffman2017quasi}%
  \BibitemOpen
  \bibfield  {author} {\bibinfo {author} {\bibfnamefont {M.~E.}\ \bibnamefont
  {Hoffman}}\ and\ \bibinfo {author} {\bibfnamefont {K.}~\bibnamefont
  {Ihara}},\ }\href {\doibase 10.1016/j.jalgebra.2017.03.005} {\bibfield
  {journal} {\bibinfo  {journal} {Journal of Algebra}\ }\textbf {\bibinfo
  {volume} {481}},\ \bibinfo {pages} {293–326} (\bibinfo {year}
  {2017})}\BibitemShut {NoStop}%
\bibitem [{\citenamefont {Fauvet}\ \emph {et~al.}(2017)\citenamefont {Fauvet},
  \citenamefont {Foissy},\ and\ \citenamefont {Manchon}}]{fauvet2017hopf}%
  \BibitemOpen
  \bibfield  {author} {\bibinfo {author} {\bibfnamefont {F.}~\bibnamefont
  {Fauvet}}, \bibinfo {author} {\bibfnamefont {L.}~\bibnamefont {Foissy}}, \
  and\ \bibinfo {author} {\bibfnamefont {D.}~\bibnamefont {Manchon}},\ }in\
  \href@noop {} {\emph {\bibinfo {booktitle} {Annales de l'Institut
  Fourier}}},\ Vol.~\bibinfo {volume} {67}\ (\bibinfo {year} {2017})\ pp.\
  \bibinfo {pages} {911--945}\BibitemShut {NoStop}%
\bibitem [{\citenamefont {Brandhuber}\ \emph
  {et~al.}(2021{\natexlab{a}})\citenamefont {Brandhuber}, \citenamefont {Chen},
  \citenamefont {Travaglini},\ and\ \citenamefont {Wen}}]{Brandhuber:2021kpo}%
  \BibitemOpen
  \bibfield  {author} {\bibinfo {author} {\bibfnamefont {A.}~\bibnamefont
  {Brandhuber}}, \bibinfo {author} {\bibfnamefont {G.}~\bibnamefont {Chen}},
  \bibinfo {author} {\bibfnamefont {G.}~\bibnamefont {Travaglini}}, \ and\
  \bibinfo {author} {\bibfnamefont {C.}~\bibnamefont {Wen}},\ }\href {\doibase
  10.1007/JHEP07(2021)047} {\bibfield  {journal} {\bibinfo  {journal} {JHEP}\
  }\textbf {\bibinfo {volume} {07}},\ \bibinfo {pages} {047} (\bibinfo {year}
  {2021}{\natexlab{a}})},\ \Eprint {http://arxiv.org/abs/2104.11206}
  {arXiv:2104.11206 [hep-th]} \BibitemShut {NoStop}%
\bibitem [{\citenamefont {Brandhuber}\ \emph
  {et~al.}(2022{\natexlab{a}})\citenamefont {Brandhuber}, \citenamefont {Chen},
  \citenamefont {Johansson}, \citenamefont {Travaglini},\ and\ \citenamefont
  {Wen}}]{Brandhuber:2021bsf}%
  \BibitemOpen
  \bibfield  {author} {\bibinfo {author} {\bibfnamefont {A.}~\bibnamefont
  {Brandhuber}}, \bibinfo {author} {\bibfnamefont {G.}~\bibnamefont {Chen}},
  \bibinfo {author} {\bibfnamefont {H.}~\bibnamefont {Johansson}}, \bibinfo
  {author} {\bibfnamefont {G.}~\bibnamefont {Travaglini}}, \ and\ \bibinfo
  {author} {\bibfnamefont {C.}~\bibnamefont {Wen}},\ }\href {\doibase
  10.1103/PhysRevLett.128.121601} {\bibfield  {journal} {\bibinfo  {journal}
  {Phys. Rev. Lett.}\ }\textbf {\bibinfo {volume} {128}},\ \bibinfo {pages}
  {121601} (\bibinfo {year} {2022}{\natexlab{a}})},\ \Eprint
  {http://arxiv.org/abs/2111.15649} {arXiv:2111.15649 [hep-th]} \BibitemShut
  {NoStop}%
\bibitem [{\citenamefont {Brandhuber}\ \emph
  {et~al.}(2022{\natexlab{b}})\citenamefont {Brandhuber}, \citenamefont
  {Brown}, \citenamefont {Chen}, \citenamefont {Gowdy}, \citenamefont
  {Travaglini},\ and\ \citenamefont {Wen}}]{Brandhuber:2022enp}%
  \BibitemOpen
  \bibfield  {author} {\bibinfo {author} {\bibfnamefont {A.}~\bibnamefont
  {Brandhuber}}, \bibinfo {author} {\bibfnamefont {G.~R.}\ \bibnamefont
  {Brown}}, \bibinfo {author} {\bibfnamefont {G.}~\bibnamefont {Chen}},
  \bibinfo {author} {\bibfnamefont {J.}~\bibnamefont {Gowdy}}, \bibinfo
  {author} {\bibfnamefont {G.}~\bibnamefont {Travaglini}}, \ and\ \bibinfo
  {author} {\bibfnamefont {C.}~\bibnamefont {Wen}},\ }\href {\doibase
  10.1007/JHEP12(2022)101} {\bibfield  {journal} {\bibinfo  {journal} {JHEP}\
  }\textbf {\bibinfo {volume} {12}},\ \bibinfo {pages} {101} (\bibinfo {year}
  {2022}{\natexlab{b}})},\ \Eprint {http://arxiv.org/abs/2208.05886}
  {arXiv:2208.05886 [hep-th]} \BibitemShut {NoStop}%
\bibitem [{\citenamefont {Chen}\ \emph {et~al.}(2023)\citenamefont {Chen},
  \citenamefont {Lin},\ and\ \citenamefont {Wen}}]{Chen:2022nei}%
  \BibitemOpen
  \bibfield  {author} {\bibinfo {author} {\bibfnamefont {G.}~\bibnamefont
  {Chen}}, \bibinfo {author} {\bibfnamefont {G.}~\bibnamefont {Lin}}, \ and\
  \bibinfo {author} {\bibfnamefont {C.}~\bibnamefont {Wen}},\ }\href {\doibase
  10.1103/PhysRevD.107.L081701} {\bibfield  {journal} {\bibinfo  {journal}
  {Phys. Rev. D}\ }\textbf {\bibinfo {volume} {107}},\ \bibinfo {pages}
  {L081701} (\bibinfo {year} {2023})},\ \Eprint
  {http://arxiv.org/abs/2208.05519} {arXiv:2208.05519 [hep-th]} \BibitemShut
  {NoStop}%
\bibitem [{\citenamefont {Lin}\ and\ \citenamefont {Yang}(2022)}]{Lin:2022jrp}%
  \BibitemOpen
  \bibfield  {author} {\bibinfo {author} {\bibfnamefont {G.}~\bibnamefont
  {Lin}}\ and\ \bibinfo {author} {\bibfnamefont {G.}~\bibnamefont {Yang}},\
  }\href@noop {} {\  (\bibinfo {year} {2022})},\ \Eprint
  {http://arxiv.org/abs/2211.01386} {arXiv:2211.01386 [hep-th]} \BibitemShut
  {NoStop}%
\bibitem [{\citenamefont {Lin}\ and\ \citenamefont {Yang}(2024)}]{Lin:2023rwe}%
  \BibitemOpen
  \bibfield  {author} {\bibinfo {author} {\bibfnamefont {G.}~\bibnamefont
  {Lin}}\ and\ \bibinfo {author} {\bibfnamefont {G.}~\bibnamefont {Yang}},\
  }\href {\doibase 10.1007/JHEP02(2024)013} {\bibfield  {journal} {\bibinfo
  {journal} {JHEP}\ }\textbf {\bibinfo {volume} {02}},\ \bibinfo {pages} {013}
  (\bibinfo {year} {2024})},\ \Eprint {http://arxiv.org/abs/2306.04672}
  {arXiv:2306.04672 [hep-th]} \BibitemShut {NoStop}%
\bibitem [{\citenamefont {Bjerrum-Bohr}\ \emph {et~al.}()\citenamefont
  {Bjerrum-Bohr}, \citenamefont {Chen}, \citenamefont {Miao},\ and\
  \citenamefont {Skowronek}}]{FermionKiHA}%
  \BibitemOpen
  \bibfield  {author} {\bibinfo {author} {\bibfnamefont {N.~E.~J.}\
  \bibnamefont {Bjerrum-Bohr}}, \bibinfo {author} {\bibfnamefont
  {G.}~\bibnamefont {Chen}}, \bibinfo {author} {\bibfnamefont {Y.}~\bibnamefont
  {Miao}}, \ and\ \bibinfo {author} {\bibfnamefont {M.}~\bibnamefont
  {Skowronek}},\ }\href@noop {} {\ }\Eprint {http://arxiv.org/abs/to appear}
  {arXiv:to appear [hep-th]} \BibitemShut {NoStop}%
\bibitem [{\citenamefont {Cao}\ \emph {et~al.}(2023)\citenamefont {Cao},
  \citenamefont {Dong}, \citenamefont {He},\ and\ \citenamefont
  {Zhang}}]{Cao:2022vou}%
  \BibitemOpen
  \bibfield  {author} {\bibinfo {author} {\bibfnamefont {Q.}~\bibnamefont
  {Cao}}, \bibinfo {author} {\bibfnamefont {J.}~\bibnamefont {Dong}}, \bibinfo
  {author} {\bibfnamefont {S.}~\bibnamefont {He}}, \ and\ \bibinfo {author}
  {\bibfnamefont {Y.-Q.}\ \bibnamefont {Zhang}},\ }\href {\doibase
  10.1103/PhysRevD.107.026022} {\bibfield  {journal} {\bibinfo  {journal}
  {Phys. Rev. D}\ }\textbf {\bibinfo {volume} {107}},\ \bibinfo {pages}
  {026022} (\bibinfo {year} {2023})},\ \Eprint
  {http://arxiv.org/abs/2211.05404} {arXiv:2211.05404 [hep-th]} \BibitemShut
  {NoStop}%
\bibitem [{\citenamefont {Chen}\ \emph {et~al.}(2024)\citenamefont {Chen},
  \citenamefont {Rodina},\ and\ \citenamefont {Wen}}]{Chen:2023ekh}%
  \BibitemOpen
  \bibfield  {author} {\bibinfo {author} {\bibfnamefont {G.}~\bibnamefont
  {Chen}}, \bibinfo {author} {\bibfnamefont {L.}~\bibnamefont {Rodina}}, \ and\
  \bibinfo {author} {\bibfnamefont {C.}~\bibnamefont {Wen}},\ }\href {\doibase
  10.1007/JHEP02(2024)096} {\bibfield  {journal} {\bibinfo  {journal} {JHEP}\
  }\textbf {\bibinfo {volume} {02}},\ \bibinfo {pages} {096} (\bibinfo {year}
  {2024})},\ \Eprint {http://arxiv.org/abs/2310.11943} {arXiv:2310.11943
  [hep-th]} \BibitemShut {NoStop}%
\bibitem [{Note1()}]{Note1}%
  \BibitemOpen
  \bibinfo {note} {See also \cite {Bonnefoy:2023imz} for a different approach
  studying the colour-kinematic duality for $\protect \rm \alpha ' F^3+\alpha
  '^2 F^4$ theory.}\BibitemShut {Stop}%
\bibitem [{\citenamefont {Johansson}\ and\ \citenamefont
  {Nohle}(2017)}]{Johansson:2017srf}%
  \BibitemOpen
  \bibfield  {author} {\bibinfo {author} {\bibfnamefont {H.}~\bibnamefont
  {Johansson}}\ and\ \bibinfo {author} {\bibfnamefont {J.}~\bibnamefont
  {Nohle}},\ }\href@noop {} {\  (\bibinfo {year} {2017})},\ \Eprint
  {http://arxiv.org/abs/1707.02965} {arXiv:1707.02965 [hep-th]} \BibitemShut
  {NoStop}%
\bibitem [{\citenamefont {Huang}\ \emph {et~al.}(2016)\citenamefont {Huang},
  \citenamefont {Schlotterer},\ and\ \citenamefont {Wen}}]{Huang:2016tag}%
  \BibitemOpen
  \bibfield  {author} {\bibinfo {author} {\bibfnamefont {Y.-t.}\ \bibnamefont
  {Huang}}, \bibinfo {author} {\bibfnamefont {O.}~\bibnamefont {Schlotterer}},
  \ and\ \bibinfo {author} {\bibfnamefont {C.}~\bibnamefont {Wen}},\ }\href
  {\doibase 10.1007/JHEP09(2016)155} {\bibfield  {journal} {\bibinfo  {journal}
  {JHEP}\ }\textbf {\bibinfo {volume} {09}},\ \bibinfo {pages} {155} (\bibinfo
  {year} {2016})},\ \Eprint {http://arxiv.org/abs/1602.01674} {arXiv:1602.01674
  [hep-th]} \BibitemShut {NoStop}%
\bibitem [{\citenamefont {Azevedo}\ \emph {et~al.}(2018)\citenamefont
  {Azevedo}, \citenamefont {Chiodaroli}, \citenamefont {Johansson},\ and\
  \citenamefont {Schlotterer}}]{Azevedo:2018dgo}%
  \BibitemOpen
  \bibfield  {author} {\bibinfo {author} {\bibfnamefont {T.}~\bibnamefont
  {Azevedo}}, \bibinfo {author} {\bibfnamefont {M.}~\bibnamefont {Chiodaroli}},
  \bibinfo {author} {\bibfnamefont {H.}~\bibnamefont {Johansson}}, \ and\
  \bibinfo {author} {\bibfnamefont {O.}~\bibnamefont {Schlotterer}},\ }\href
  {\doibase 10.1007/JHEP10(2018)012} {\bibfield  {journal} {\bibinfo  {journal}
  {JHEP}\ }\textbf {\bibinfo {volume} {10}},\ \bibinfo {pages} {012} (\bibinfo
  {year} {2018})},\ \Eprint {http://arxiv.org/abs/1803.05452} {arXiv:1803.05452
  [hep-th]} \BibitemShut {NoStop}%
\bibitem [{\citenamefont {Kawai}\ \emph {et~al.}(1986)\citenamefont {Kawai},
  \citenamefont {Lewellen},\ and\ \citenamefont {Tye}}]{Kawai:1985xq}%
  \BibitemOpen
  \bibfield  {author} {\bibinfo {author} {\bibfnamefont {H.}~\bibnamefont
  {Kawai}}, \bibinfo {author} {\bibfnamefont {D.~C.}\ \bibnamefont {Lewellen}},
  \ and\ \bibinfo {author} {\bibfnamefont {S.~H.~H.}\ \bibnamefont {Tye}},\
  }\href {\doibase 10.1016/0550-3213(86)90362-7} {\bibfield  {journal}
  {\bibinfo  {journal} {Nucl. Phys.}\ }\textbf {\bibinfo {volume} {B269}},\
  \bibinfo {pages} {1} (\bibinfo {year} {1986})}\BibitemShut {NoStop}%
\bibitem [{\citenamefont {Cheung}\ and\ \citenamefont
  {Mangan}(2021)}]{Cheung:2021zvb}%
  \BibitemOpen
  \bibfield  {author} {\bibinfo {author} {\bibfnamefont {C.}~\bibnamefont
  {Cheung}}\ and\ \bibinfo {author} {\bibfnamefont {J.}~\bibnamefont
  {Mangan}},\ }\href {\doibase 10.1007/JHEP11(2021)069} {\bibfield  {journal}
  {\bibinfo  {journal} {JHEP}\ }\textbf {\bibinfo {volume} {11}},\ \bibinfo
  {pages} {069} (\bibinfo {year} {2021})},\ \Eprint
  {http://arxiv.org/abs/2108.02276} {arXiv:2108.02276 [hep-th]} \BibitemShut
  {NoStop}%
\bibitem [{\citenamefont {Vaman}\ and\ \citenamefont
  {Yao}(2010)}]{Vaman:2010ez}%
  \BibitemOpen
  \bibfield  {author} {\bibinfo {author} {\bibfnamefont {D.}~\bibnamefont
  {Vaman}}\ and\ \bibinfo {author} {\bibfnamefont {Y.-P.}\ \bibnamefont
  {Yao}},\ }\href {\doibase 10.1007/JHEP11(2010)028} {\bibfield  {journal}
  {\bibinfo  {journal} {JHEP}\ }\textbf {\bibinfo {volume} {11}},\ \bibinfo
  {pages} {028} (\bibinfo {year} {2010})},\ \Eprint
  {http://arxiv.org/abs/1007.3475} {arXiv:1007.3475 [hep-th]} \BibitemShut
  {NoStop}%
\bibitem [{\citenamefont {Azevedo}\ and\ \citenamefont
  {Engelund}(2017)}]{Azevedo:2017lkz}%
  \BibitemOpen
  \bibfield  {author} {\bibinfo {author} {\bibfnamefont {T.}~\bibnamefont
  {Azevedo}}\ and\ \bibinfo {author} {\bibfnamefont {O.~T.}\ \bibnamefont
  {Engelund}},\ }\href {\doibase 10.1007/JHEP11(2017)052} {\bibfield  {journal}
  {\bibinfo  {journal} {JHEP}\ }\textbf {\bibinfo {volume} {11}},\ \bibinfo
  {pages} {052} (\bibinfo {year} {2017})},\ \Eprint
  {http://arxiv.org/abs/1707.02192} {arXiv:1707.02192 [hep-th]} \BibitemShut
  {NoStop}%
\bibitem [{\citenamefont {Garozzo}\ \emph {et~al.}(2019)\citenamefont
  {Garozzo}, \citenamefont {Queimada},\ and\ \citenamefont
  {Schlotterer}}]{Garozzo:2018uzj}%
  \BibitemOpen
  \bibfield  {author} {\bibinfo {author} {\bibfnamefont {L.~M.}\ \bibnamefont
  {Garozzo}}, \bibinfo {author} {\bibfnamefont {L.}~\bibnamefont {Queimada}}, \
  and\ \bibinfo {author} {\bibfnamefont {O.}~\bibnamefont {Schlotterer}},\
  }\href {\doibase 10.1007/JHEP02(2019)078} {\bibfield  {journal} {\bibinfo
  {journal} {JHEP}\ }\textbf {\bibinfo {volume} {02}},\ \bibinfo {pages} {078}
  (\bibinfo {year} {2019})},\ \Eprint {http://arxiv.org/abs/1809.08103}
  {arXiv:1809.08103 [hep-th]} \BibitemShut {NoStop}%
\bibitem [{\citenamefont {Menezes}(2022)}]{Menezes:2021dyp}%
  \BibitemOpen
  \bibfield  {author} {\bibinfo {author} {\bibfnamefont {G.}~\bibnamefont
  {Menezes}},\ }\href {\doibase 10.1007/JHEP03(2022)074} {\bibfield  {journal}
  {\bibinfo  {journal} {JHEP}\ }\textbf {\bibinfo {volume} {03}},\ \bibinfo
  {pages} {074} (\bibinfo {year} {2022})},\ \Eprint
  {http://arxiv.org/abs/2112.00978} {arXiv:2112.00978 [hep-th]} \BibitemShut
  {NoStop}%
\bibitem [{\citenamefont {Carrasco}\ \emph
  {et~al.}(2023{\natexlab{a}})\citenamefont {Carrasco}, \citenamefont
  {Lewandowski},\ and\ \citenamefont {Pavao}}]{Carrasco:2022sck}%
  \BibitemOpen
  \bibfield  {author} {\bibinfo {author} {\bibfnamefont {J.~J.~M.}\
  \bibnamefont {Carrasco}}, \bibinfo {author} {\bibfnamefont {M.}~\bibnamefont
  {Lewandowski}}, \ and\ \bibinfo {author} {\bibfnamefont {N.~H.}\ \bibnamefont
  {Pavao}},\ }\href {\doibase 10.1007/JHEP02(2023)015} {\bibfield  {journal}
  {\bibinfo  {journal} {JHEP}\ }\textbf {\bibinfo {volume} {02}},\ \bibinfo
  {pages} {015} (\bibinfo {year} {2023}{\natexlab{a}})},\ \Eprint
  {http://arxiv.org/abs/2211.04441} {arXiv:2211.04441 [hep-th]} \BibitemShut
  {NoStop}%
\bibitem [{\citenamefont {Carrasco}\ \emph
  {et~al.}(2023{\natexlab{b}})\citenamefont {Carrasco}, \citenamefont
  {Lewandowski},\ and\ \citenamefont {Pavao}}]{Carrasco:2022lbm}%
  \BibitemOpen
  \bibfield  {author} {\bibinfo {author} {\bibfnamefont {J.~J.~M.}\
  \bibnamefont {Carrasco}}, \bibinfo {author} {\bibfnamefont {M.}~\bibnamefont
  {Lewandowski}}, \ and\ \bibinfo {author} {\bibfnamefont {N.~H.}\ \bibnamefont
  {Pavao}},\ }\href {\doibase 10.1103/PhysRevLett.131.051601} {\bibfield
  {journal} {\bibinfo  {journal} {Phys. Rev. Lett.}\ }\textbf {\bibinfo
  {volume} {131}},\ \bibinfo {pages} {051601} (\bibinfo {year}
  {2023}{\natexlab{b}})},\ \Eprint {http://arxiv.org/abs/2203.03592}
  {arXiv:2203.03592 [hep-th]} \BibitemShut {NoStop}%
\bibitem [{\citenamefont {Carrasco}\ and\ \citenamefont
  {Pavao}(2023)}]{Carrasco:2023wib}%
  \BibitemOpen
  \bibfield  {author} {\bibinfo {author} {\bibfnamefont {J.~J.~M.}\
  \bibnamefont {Carrasco}}\ and\ \bibinfo {author} {\bibfnamefont {N.~H.}\
  \bibnamefont {Pavao}},\ }\href@noop {} {\  (\bibinfo {year} {2023})},\
  \Eprint {http://arxiv.org/abs/2310.06316} {arXiv:2310.06316 [hep-ph]}
  \BibitemShut {NoStop}%
\bibitem [{\citenamefont {Bonnefoy}\ \emph {et~al.}(2023)\citenamefont
  {Bonnefoy}, \citenamefont {Durieux},\ and\ \citenamefont
  {Roosmale~Nepveu}}]{Bonnefoy:2023imz}%
  \BibitemOpen
  \bibfield  {author} {\bibinfo {author} {\bibfnamefont {Q.}~\bibnamefont
  {Bonnefoy}}, \bibinfo {author} {\bibfnamefont {G.}~\bibnamefont {Durieux}}, \
  and\ \bibinfo {author} {\bibfnamefont {J.}~\bibnamefont {Roosmale~Nepveu}},\
  }\href@noop {} {\  (\bibinfo {year} {2023})},\ \Eprint
  {http://arxiv.org/abs/2310.13041} {arXiv:2310.13041 [hep-th]} \BibitemShut
  {NoStop}%
\bibitem [{\citenamefont {Garozzo}\ and\ \citenamefont
  {Guevara}(2024)}]{Garozzo:2024myw}%
  \BibitemOpen
  \bibfield  {author} {\bibinfo {author} {\bibfnamefont {L.}~\bibnamefont
  {Garozzo}}\ and\ \bibinfo {author} {\bibfnamefont {A.}~\bibnamefont
  {Guevara}},\ }\href@noop {} {\  (\bibinfo {year} {2024})},\ \Eprint
  {http://arxiv.org/abs/2402.19430} {arXiv:2402.19430 [hep-th]} \BibitemShut
  {NoStop}%
\bibitem [{\citenamefont {Azevedo}\ \emph {et~al.}(2020)\citenamefont
  {Azevedo}, \citenamefont {Jusinskas},\ and\ \citenamefont
  {Lize}}]{Azevedo:2019zbn}%
  \BibitemOpen
  \bibfield  {author} {\bibinfo {author} {\bibfnamefont {T.}~\bibnamefont
  {Azevedo}}, \bibinfo {author} {\bibfnamefont {R.~L.}\ \bibnamefont
  {Jusinskas}}, \ and\ \bibinfo {author} {\bibfnamefont {M.}~\bibnamefont
  {Lize}},\ }\href {\doibase 10.1007/JHEP01(2020)082} {\bibfield  {journal}
  {\bibinfo  {journal} {JHEP}\ }\textbf {\bibinfo {volume} {01}},\ \bibinfo
  {pages} {082} (\bibinfo {year} {2020})},\ \Eprint
  {http://arxiv.org/abs/1908.11371} {arXiv:1908.11371 [hep-th]} \BibitemShut
  {NoStop}%
\bibitem [{Note2()}]{Note2}%
  \BibitemOpen
  \bibinfo {note} {Due to the cyclic summation in the operator $\protect
  \mathbb {O}$, if ${\protect \rm cyc}( x_1,\protect \ldots , x_s)= {\protect
  \rm cyc}( x'_1,\protect \ldots , x'_s)$, the two partitions generate
  identical terms under the action of $\protect \mathbb {O}_{{\protect \rm
  cyc}([i_1\protect \ldots i_{r-1}]i_r)}$. Therefore, for each partition
  element with $s$ subsets, the $\protect \mathbb {O}$ operator introduces an
  overcounting factor equal to $s$.}\BibitemShut {Stop}%
\bibitem [{\citenamefont {Du}\ and\ \citenamefont {Teng}(2017)}]{Du:2017kpo}%
  \BibitemOpen
  \bibfield  {author} {\bibinfo {author} {\bibfnamefont {Y.-J.}\ \bibnamefont
  {Du}}\ and\ \bibinfo {author} {\bibfnamefont {F.}~\bibnamefont {Teng}},\
  }\href {\doibase 10.1007/JHEP04(2017)033} {\bibfield  {journal} {\bibinfo
  {journal} {JHEP}\ }\textbf {\bibinfo {volume} {04}},\ \bibinfo {pages} {033}
  (\bibinfo {year} {2017})},\ \Eprint {http://arxiv.org/abs/1703.05717}
  {arXiv:1703.05717 [hep-th]} \BibitemShut {NoStop}%
\bibitem [{\citenamefont {Edison}\ and\ \citenamefont
  {Teng}(2020)}]{Edison:2020ehu}%
  \BibitemOpen
  \bibfield  {author} {\bibinfo {author} {\bibfnamefont {A.}~\bibnamefont
  {Edison}}\ and\ \bibinfo {author} {\bibfnamefont {F.}~\bibnamefont {Teng}},\
  }\href {\doibase 10.1007/JHEP12(2020)138} {\bibfield  {journal} {\bibinfo
  {journal} {JHEP}\ }\textbf {\bibinfo {volume} {12}},\ \bibinfo {pages} {138}
  (\bibinfo {year} {2020})},\ \Eprint {http://arxiv.org/abs/2005.03638}
  {arXiv:2005.03638 [hep-th]} \BibitemShut {NoStop}%
\bibitem [{\citenamefont {Chen}(2022)}]{ChenGitHub}%
  \BibitemOpen
  \bibfield  {author} {\bibinfo {author} {\bibfnamefont {G.}~\bibnamefont
  {Chen}},\ }\href {https://github.com/AmplitudeGravity/kinematicHopfAlgebra}
  {\enquote {\bibinfo {title}
  {https://github.com/amplitudegravity/\\kinematichopfalgebra},}\ } (\bibinfo
  {year} {2022})\BibitemShut {NoStop}%
\bibitem [{\citenamefont {Gross}(1962)}]{10.2307/2312725}%
  \BibitemOpen
  \bibfield  {author} {\bibinfo {author} {\bibfnamefont {O.~A.}\ \bibnamefont
  {Gross}},\ }\href {http://www.jstor.org/stable/2312725} {\bibfield  {journal}
  {\bibinfo  {journal} {The American Mathematical Monthly}\ }\textbf {\bibinfo
  {volume} {69}},\ \bibinfo {pages} {4} (\bibinfo {year} {1962})}\BibitemShut
  {NoStop}%
\bibitem [{\citenamefont {Cheung}\ \emph {et~al.}(2018)\citenamefont {Cheung},
  \citenamefont {Shen},\ and\ \citenamefont {Wen}}]{Cheung:2017ems}%
  \BibitemOpen
  \bibfield  {author} {\bibinfo {author} {\bibfnamefont {C.}~\bibnamefont
  {Cheung}}, \bibinfo {author} {\bibfnamefont {C.-H.}\ \bibnamefont {Shen}}, \
  and\ \bibinfo {author} {\bibfnamefont {C.}~\bibnamefont {Wen}},\ }\href
  {\doibase 10.1007/JHEP02(2018)095} {\bibfield  {journal} {\bibinfo  {journal}
  {JHEP}\ }\textbf {\bibinfo {volume} {02}},\ \bibinfo {pages} {095} (\bibinfo
  {year} {2018})},\ \Eprint {http://arxiv.org/abs/1705.03025} {arXiv:1705.03025
  [hep-th]} \BibitemShut {NoStop}%
\bibitem [{\citenamefont {Chiodaroli}\ \emph {et~al.}(2017)\citenamefont
  {Chiodaroli}, \citenamefont {Gunaydin}, \citenamefont {Johansson},\ and\
  \citenamefont {Roiban}}]{Chiodaroli:2015rdg}%
  \BibitemOpen
  \bibfield  {author} {\bibinfo {author} {\bibfnamefont {M.}~\bibnamefont
  {Chiodaroli}}, \bibinfo {author} {\bibfnamefont {M.}~\bibnamefont
  {Gunaydin}}, \bibinfo {author} {\bibfnamefont {H.}~\bibnamefont {Johansson}},
  \ and\ \bibinfo {author} {\bibfnamefont {R.}~\bibnamefont {Roiban}},\ }\href
  {\doibase 10.1007/JHEP06(2017)064} {\bibfield  {journal} {\bibinfo  {journal}
  {JHEP}\ }\textbf {\bibinfo {volume} {06}},\ \bibinfo {pages} {064} (\bibinfo
  {year} {2017})},\ \Eprint {http://arxiv.org/abs/1511.01740} {arXiv:1511.01740
  [hep-th]} \BibitemShut {NoStop}%
\bibitem [{\citenamefont {Carrasco}\ \emph {et~al.}(2020)\citenamefont
  {Carrasco}, \citenamefont {Rodina}, \citenamefont {Yin},\ and\ \citenamefont
  {Zekioglu}}]{Carrasco:2019yyn}%
  \BibitemOpen
  \bibfield  {author} {\bibinfo {author} {\bibfnamefont {J.~J.~M.}\
  \bibnamefont {Carrasco}}, \bibinfo {author} {\bibfnamefont {L.}~\bibnamefont
  {Rodina}}, \bibinfo {author} {\bibfnamefont {Z.}~\bibnamefont {Yin}}, \ and\
  \bibinfo {author} {\bibfnamefont {S.}~\bibnamefont {Zekioglu}},\ }\href
  {\doibase 10.1103/PhysRevLett.125.251602} {\bibfield  {journal} {\bibinfo
  {journal} {Phys. Rev. Lett.}\ }\textbf {\bibinfo {volume} {125}},\ \bibinfo
  {pages} {251602} (\bibinfo {year} {2020})},\ \Eprint
  {http://arxiv.org/abs/1910.12850} {arXiv:1910.12850 [hep-th]} \BibitemShut
  {NoStop}%
\bibitem [{\citenamefont {Low}\ \emph {et~al.}(2021)\citenamefont {Low},
  \citenamefont {Rodina},\ and\ \citenamefont {Yin}}]{Low:2020ubn}%
  \BibitemOpen
  \bibfield  {author} {\bibinfo {author} {\bibfnamefont {I.}~\bibnamefont
  {Low}}, \bibinfo {author} {\bibfnamefont {L.}~\bibnamefont {Rodina}}, \ and\
  \bibinfo {author} {\bibfnamefont {Z.}~\bibnamefont {Yin}},\ }\href {\doibase
  10.1103/PhysRevD.103.025004} {\bibfield  {journal} {\bibinfo  {journal}
  {Phys. Rev. D}\ }\textbf {\bibinfo {volume} {103}},\ \bibinfo {pages}
  {025004} (\bibinfo {year} {2021})},\ \Eprint
  {http://arxiv.org/abs/2009.00008} {arXiv:2009.00008 [hep-th]} \BibitemShut
  {NoStop}%
\bibitem [{\citenamefont {Carrasco}\ \emph {et~al.}(2021)\citenamefont
  {Carrasco}, \citenamefont {Rodina},\ and\ \citenamefont
  {Zekioglu}}]{Carrasco:2021ptp}%
  \BibitemOpen
  \bibfield  {author} {\bibinfo {author} {\bibfnamefont {J.~J.~M.}\
  \bibnamefont {Carrasco}}, \bibinfo {author} {\bibfnamefont {L.}~\bibnamefont
  {Rodina}}, \ and\ \bibinfo {author} {\bibfnamefont {S.}~\bibnamefont
  {Zekioglu}},\ }\href {\doibase 10.1007/JHEP06(2021)169} {\bibfield  {journal}
  {\bibinfo  {journal} {JHEP}\ }\textbf {\bibinfo {volume} {06}},\ \bibinfo
  {pages} {169} (\bibinfo {year} {2021})},\ \Eprint
  {http://arxiv.org/abs/2104.08370} {arXiv:2104.08370 [hep-th]} \BibitemShut
  {NoStop}%
\bibitem [{\citenamefont {Pavao}(2023)}]{Pavao:2022kog}%
  \BibitemOpen
  \bibfield  {author} {\bibinfo {author} {\bibfnamefont {N.~H.}\ \bibnamefont
  {Pavao}},\ }\href {\doibase 10.1103/PhysRevD.107.065020} {\bibfield
  {journal} {\bibinfo  {journal} {Phys. Rev. D}\ }\textbf {\bibinfo {volume}
  {107}},\ \bibinfo {pages} {065020} (\bibinfo {year} {2023})},\ \Eprint
  {http://arxiv.org/abs/2210.12800} {arXiv:2210.12800 [hep-th]} \BibitemShut
  {NoStop}%
\bibitem [{\citenamefont {Brandhuber}\ and\ \citenamefont
  {Travaglini}(2020)}]{Brandhuber:2019qpg}%
  \BibitemOpen
  \bibfield  {author} {\bibinfo {author} {\bibfnamefont {A.}~\bibnamefont
  {Brandhuber}}\ and\ \bibinfo {author} {\bibfnamefont {G.}~\bibnamefont
  {Travaglini}},\ }\href {\doibase 10.1007/JHEP01(2020)010} {\bibfield
  {journal} {\bibinfo  {journal} {JHEP}\ }\textbf {\bibinfo {volume} {01}},\
  \bibinfo {pages} {010} (\bibinfo {year} {2020})},\ \Eprint
  {http://arxiv.org/abs/1905.05657} {arXiv:1905.05657 [hep-th]} \BibitemShut
  {NoStop}%
\bibitem [{\citenamefont {Accettulli~Huber}\ \emph {et~al.}(2020)\citenamefont
  {Accettulli~Huber}, \citenamefont {Brandhuber}, \citenamefont {De~Angelis},\
  and\ \citenamefont {Travaglini}}]{AccettulliHuber:2020oou}%
  \BibitemOpen
  \bibfield  {author} {\bibinfo {author} {\bibfnamefont {M.}~\bibnamefont
  {Accettulli~Huber}}, \bibinfo {author} {\bibfnamefont {A.}~\bibnamefont
  {Brandhuber}}, \bibinfo {author} {\bibfnamefont {S.}~\bibnamefont
  {De~Angelis}}, \ and\ \bibinfo {author} {\bibfnamefont {G.}~\bibnamefont
  {Travaglini}},\ }\href {\doibase 10.1103/PhysRevD.102.046014} {\bibfield
  {journal} {\bibinfo  {journal} {Phys. Rev. D}\ }\textbf {\bibinfo {volume}
  {102}},\ \bibinfo {pages} {046014} (\bibinfo {year} {2020})},\ \Eprint
  {http://arxiv.org/abs/2006.02375} {arXiv:2006.02375 [hep-th]} \BibitemShut
  {NoStop}%
\bibitem [{\citenamefont {Accettulli~Huber}\ \emph {et~al.}(2021)\citenamefont
  {Accettulli~Huber}, \citenamefont {Brandhuber}, \citenamefont {De~Angelis},\
  and\ \citenamefont {Travaglini}}]{AccettulliHuber:2020dal}%
  \BibitemOpen
  \bibfield  {author} {\bibinfo {author} {\bibfnamefont {M.}~\bibnamefont
  {Accettulli~Huber}}, \bibinfo {author} {\bibfnamefont {A.}~\bibnamefont
  {Brandhuber}}, \bibinfo {author} {\bibfnamefont {S.}~\bibnamefont
  {De~Angelis}}, \ and\ \bibinfo {author} {\bibfnamefont {G.}~\bibnamefont
  {Travaglini}},\ }\href {\doibase 10.1103/PhysRevD.103.045015} {\bibfield
  {journal} {\bibinfo  {journal} {Phys. Rev. D}\ }\textbf {\bibinfo {volume}
  {103}},\ \bibinfo {pages} {045015} (\bibinfo {year} {2021})},\ \Eprint
  {http://arxiv.org/abs/2012.06548} {arXiv:2012.06548 [hep-th]} \BibitemShut
  {NoStop}%
\bibitem [{\citenamefont {Emond}\ and\ \citenamefont
  {Moynihan}(2019)}]{Emond:2019crr}%
  \BibitemOpen
  \bibfield  {author} {\bibinfo {author} {\bibfnamefont {W.~T.}\ \bibnamefont
  {Emond}}\ and\ \bibinfo {author} {\bibfnamefont {N.}~\bibnamefont
  {Moynihan}},\ }\href {\doibase 10.1007/JHEP12(2019)019} {\bibfield  {journal}
  {\bibinfo  {journal} {JHEP}\ }\textbf {\bibinfo {volume} {12}},\ \bibinfo
  {pages} {019} (\bibinfo {year} {2019})},\ \Eprint
  {http://arxiv.org/abs/1905.08213} {arXiv:1905.08213 [hep-th]} \BibitemShut
  {NoStop}%
\bibitem [{\citenamefont {Sennett}\ \emph {et~al.}(2019)\citenamefont
  {Sennett}, \citenamefont {Brito}, \citenamefont {Buonanno}, \citenamefont
  {Gorbenko},\ and\ \citenamefont {Senatore}}]{Sennett:2019bpc}%
  \BibitemOpen
  \bibfield  {author} {\bibinfo {author} {\bibfnamefont {N.}~\bibnamefont
  {Sennett}}, \bibinfo {author} {\bibfnamefont {R.}~\bibnamefont {Brito}},
  \bibinfo {author} {\bibfnamefont {A.}~\bibnamefont {Buonanno}}, \bibinfo
  {author} {\bibfnamefont {V.}~\bibnamefont {Gorbenko}}, \ and\ \bibinfo
  {author} {\bibfnamefont {L.}~\bibnamefont {Senatore}},\ }\href@noop {} {\
  (\bibinfo {year} {2019})},\ \Eprint {http://arxiv.org/abs/1912.09917}
  {arXiv:1912.09917 [gr-qc]} \BibitemShut {NoStop}%
\bibitem [{\citenamefont {Brandhuber}\ \emph
  {et~al.}(2021{\natexlab{b}})\citenamefont {Brandhuber}, \citenamefont {Chen},
  \citenamefont {Travaglini},\ and\ \citenamefont {Wen}}]{Brandhuber:2021eyq}%
  \BibitemOpen
  \bibfield  {author} {\bibinfo {author} {\bibfnamefont {A.}~\bibnamefont
  {Brandhuber}}, \bibinfo {author} {\bibfnamefont {G.}~\bibnamefont {Chen}},
  \bibinfo {author} {\bibfnamefont {G.}~\bibnamefont {Travaglini}}, \ and\
  \bibinfo {author} {\bibfnamefont {C.}~\bibnamefont {Wen}},\ }\href {\doibase
  10.1007/JHEP10(2021)118} {\bibfield  {journal} {\bibinfo  {journal} {JHEP}\
  }\textbf {\bibinfo {volume} {10}},\ \bibinfo {pages} {118} (\bibinfo {year}
  {2021}{\natexlab{b}})},\ \Eprint {http://arxiv.org/abs/2108.04216}
  {arXiv:2108.04216 [hep-th]} \BibitemShut {NoStop}%
\bibitem [{\citenamefont {Brandhuber}\ \emph {et~al.}(2023)\citenamefont
  {Brandhuber}, \citenamefont {Brown}, \citenamefont {Chen}, \citenamefont
  {De~Angelis}, \citenamefont {Gowdy},\ and\ \citenamefont
  {Travaglini}}]{Brandhuber:2023hhy}%
  \BibitemOpen
  \bibfield  {author} {\bibinfo {author} {\bibfnamefont {A.}~\bibnamefont
  {Brandhuber}}, \bibinfo {author} {\bibfnamefont {G.~R.}\ \bibnamefont
  {Brown}}, \bibinfo {author} {\bibfnamefont {G.}~\bibnamefont {Chen}},
  \bibinfo {author} {\bibfnamefont {S.}~\bibnamefont {De~Angelis}}, \bibinfo
  {author} {\bibfnamefont {J.}~\bibnamefont {Gowdy}}, \ and\ \bibinfo {author}
  {\bibfnamefont {G.}~\bibnamefont {Travaglini}},\ }\href {\doibase
  10.1007/JHEP06(2023)048} {\bibfield  {journal} {\bibinfo  {journal} {JHEP}\
  }\textbf {\bibinfo {volume} {06}},\ \bibinfo {pages} {048} (\bibinfo {year}
  {2023})},\ \Eprint {http://arxiv.org/abs/2303.06111} {arXiv:2303.06111
  [hep-th]} \BibitemShut {NoStop}%
\bibitem [{\citenamefont {Herderschee}\ \emph {et~al.}(2023)\citenamefont
  {Herderschee}, \citenamefont {Roiban},\ and\ \citenamefont
  {Teng}}]{Herderschee:2023fxh}%
  \BibitemOpen
  \bibfield  {author} {\bibinfo {author} {\bibfnamefont {A.}~\bibnamefont
  {Herderschee}}, \bibinfo {author} {\bibfnamefont {R.}~\bibnamefont {Roiban}},
  \ and\ \bibinfo {author} {\bibfnamefont {F.}~\bibnamefont {Teng}},\ }\href
  {\doibase 10.1007/JHEP06(2023)004} {\bibfield  {journal} {\bibinfo  {journal}
  {JHEP}\ }\textbf {\bibinfo {volume} {06}},\ \bibinfo {pages} {004} (\bibinfo
  {year} {2023})},\ \Eprint {http://arxiv.org/abs/2303.06112} {arXiv:2303.06112
  [hep-th]} \BibitemShut {NoStop}%
\end{thebibliography}%
\onecolumngrid

\newpage

\setcounter{secnumdepth}{2}

\appendix
\section{Minimal introduction to kinematic Hopf algebra}\label{app1}
We start from constructing a combinatorial algebra numerator obtained from the fusion product of algebraic  generators  $T_{(i)}$ as follows, 
\begin{align} \label{eq:Nhatapp}
    \widehat\npre(12\ldots n{-}2, v)=T_{(1)}\star T_{(2)}\cdots \star T_{(n{-}2)}\, ,
\end{align}
with the fusion product defined as the standard quasi-shuffle product. 
For instance, for the cases with two, three, and four gluons, we have 
\begin{align}
    \widehat\npre(12, v)=&\,\, T_{(1)}\star T_{(2)}=-T_{(12)}\, , \\
    \widehat \npre(123, v)=\,\, & T_{(1)}\star T_{(2)}\star T_{(3)}=T_{(123)}-T_{(12),(3)}-T_{(13),(2)}\,, \\
    \widehat   \npre(1234, v)=&-T_{(1234)}+T_{(123),(4)}+T_{(14),(23)}+T_{(124),(3)}+T_{(12),(34)}-T_{(12),(3),(4)}\nn\\
      &-T_{(12),(4),(3)} -T_{(14),(2),(3)} +T_{(134),(2)}+T_{(13),(24)}-T_{(13),(2),(4)}\nn\\
      &-T_{(13),(4),(2)}-T_{(14),(3),(2)} \, .
\end{align}
In general, the fusion product is captured by the following formula, 
\begin{align}
\label{fusionone}
	 &T_{(1\tau_1),\ldots, (\tau_r )}\star T_{(j)}= \hskip-0.5cm  \sum_{\sigma\in \{(\tau_1),\ldots, (\tau_{r})\}\shuffle \{(j)\}}T_{(1\sigma_1),\ldots, (\sigma_{r+1})} -\sum_{i=1}^{r}T_{(1\tau_1),\ldots, (\tau_{i-1}),(\tau_{i}j),(\tau_{i+1}), \ldots,(\tau_{r})}\, .
\end{align}

The evaluation map from generators to kinematic functions is defined as 
\begin{align}\label{eq:T}
&\la T_{(1),\ldots}\ra =0 \, ,  \qquad \la T_{\ldots,(1\ldots),\ldots}\ra=0 \,\\
&\langle T_{(1\tau_1),(\tau_2),\ldots,(\tau_r)}\rangle:=\begin{tikzpicture}[baseline={([yshift=-0.8ex]current bounding box.center)}]\tikzstyle{every node}=[font=\small]   
   \begin{feynman}
    \vertex (a)[myblob]{};
     \vertex[left=0.8cm of a] (a0)[myblob]{};
     \vertex[right=0.8cm of a] (a2)[myblob]{};
      \vertex[right=0.8cm of a2] (a3)[myblob]{};
       \vertex[right=0.8cm of a3] (a4)[myblob]{};
       \vertex[above=0.8cm of a] (b1){$\tau_1~~~~$};
        \vertex[above=0.8cm of a2] (b2){$\tau_2$};
        \vertex[above=0.8cm of a3] (b3){$\cdots$};
         \vertex[above=0.8cm of a4] (b4){$\tau_r$};
         \vertex[above=0.8cm of a0] (b0){$1~~$};
       \vertex [above=0.8cm of a0](j1){$ $};
    \vertex [right=0.2cm of j1](j2){$ $};
    \vertex [right=0.6cm of j2](j3){$ $};
    \vertex [right=0.4cm of j3](j4){$ $};
    \vertex [right=0.8cm of j4](j5){$ $};
      \vertex [right=0.2cm of j5](j6){$ $};
    \vertex [right=0.6cm of j6](j7){$ $};
     \vertex [right=0.0cm of j7](j8){$ $};
    \vertex [right=0.8cm of j8](j9){$ $};
   	 \diagram*{(a)--[very thick](a0),(a)--[very thick](a2),(a2)--[very thick](a3), (a3)--[very thick](a4),(a0) -- [thick] (j1),(a) -- [thick] (j2),(a)--[thick](j3),(a2) -- [thick] (j4),(a2)--[thick](j5),(a4) -- [thick] (j8),(a4)--[thick](j9)};
    \end{feynman}  
  \end{tikzpicture}={1\over n-2} {G_{1\tau_1}(v)\over v\mdot p_{1}}  {G_{\tau_2}(p_{\Theta(\tau_{2})})\over v\mdot p_{1\tau_1}}\, \cdots {G_{\tau_r}(p_{\Theta(\tau_{r})})\over v\mdot p_{1\tau_1\ldots \tau_{r-1}}} \, ,\nn
\end{align}
where the external lines are gluons, grouped into sets denoted as $1, \tau_1, \tau_2, \ldots, \tau_r$, which are  partitions of 
 $\{1,2,3, \ldots ,  n{-}2\}$.  

\section{From BCJ numerators to amplitude with one massive particle}\label{app2}


Here we construct the amplitude with one massive gluon or a tachyon from the corresponding BCJ numerators.  We begin with the amplitude with one transverse massive gluon. It is given by two parts. The first contribution arises from the usual trivalent graphs, 
  \begin{align} \label{eq:Am1}
  	A_{n-1}(1\ldots n-2, (n{-}1)_{g'}) \big{\vert}_{\rm part \, 1}
  &= \sum_{\Gamma}{\npre(\Gamma,(n-1)_{g'})\over d_{\Gamma}}\, ,
  \end{align}
  where $(n{-}1)_{g'}$ is the massive gluon and $d_{\Gamma}$ denotes the massless propagators in the cubic graph $\Gamma$ and the summation is over all the cubic graphs of the same color ordering. For the amplitude with a massive gluon, we also need take into account the second contribution from following diagram 
   \begin{align}
 	\begin{tikzpicture}[baseline={([yshift=-0.8ex]current bounding box.center)}]\, \tikzstyle{every node}=[font=\small]    
   \begin{feynman}
    \vertex (a)[]{$(n-1)_{g'}$};
     \vertex [above=0.9cm of a](b)[dot]{};
     \vertex [left=0.9cm of b](c);
     \vertex [left=0.42cm of b](cL);
     \vertex [above=0.33cm of cL](vL)[HV]{\tiny$\Gamma_1$};
     \vertex [right=0.83cm of vL](vR)[HV]{\tiny$\Gamma_2$};
     \vertex [above=0.6cm of vL](t1){$\sigma_{1}$};
     \vertex [above=0.6cm of vR](t2){$~~\sigma_{2}$};
    \vertex [above=1.1cm of c](j1){};
    \vertex [right=.8cm of j1](j2){};
     \vertex [right=.2cm of j2](j3){};
    \vertex [right=0.8cm of j3](j4){};
   	 \diagram*{(a) -- [thick] (b),(b)--[thick](vL) -- [thick] (j1),(vL) -- [thick] (j2),(vR)--[thick](j3),(b)--[thick](vR)--[thick](j4)};
    \end{feynman}
  \end{tikzpicture} \, ,
   \end{align}
   which gives to, 
\begin{align} \label{eq:Am2}
      	A_{n-1}(1\ldots n-2, (n{-}1)_{g'}) \big{\vert}_{\rm part \, 2} = = \sum_{\Gamma_1,\Gamma_2}{ (-1)^{n-3}W'(Z_{(1,\Gamma_1)},Z_{(\Gamma_{2},n-2)})\over d_{\Gamma_1}d_{\Gamma_2}}{-p_{\sigma_1\sigma_2}\mdot \varepsilon_{n-1}\over 2}\, ,
\end{align}
 where $\Gamma_1$ represents all cubic graphs that include gluon 1 as their first leg, and $\Gamma_2$ denotes all cubic graphs with gluon $n-2$ as their last leg.  $Z_{(1,\Gamma_1)}$ denotes all the  gluon index sequences in the left nested commutator with 1 in the first position, e.g. $Z_{(1,[[1,2],[3,4]])}=(1234)-(1243)$. $Z_{(\Gamma_2,n-2)}$ denotes all the  gluon index sequences in the left nested commutator $\Gamma_2$ with $n-2$ in the last position. Then the final amplitude is given by the sum of \eqref{eq:Am1} and \eqref{eq:Am2}. 
 
 We can break the graph $\Gamma$ in \eqref{eq:Am1} into sum of smaller graphs in the same form as \eqref{eq:Am2}, and the final expression of the amplitude is then given as
  \begin{align}
  &	A_{n-1}(1\ldots n-2, (n{-}1)_{g'})
  = \left({1\over\alpha'}- p_{12\ldots n-2}^2 \right)\times\left[\sum_{\Gamma_1,\Gamma_2}{ (-1)^{n-3}W'(Z_{(1,\Gamma_1)},Z_{(\Gamma_{2},n-2)})\over d_{\Gamma_1}d_{\Gamma_2}}{(p_{\sigma_2}-p_{\sigma_1})\mdot \varepsilon_{n-1}\over 2}\right]\, ,
    \end{align}
where we have used \eqref{eq:WpWp} to express the BCJ numerator in \eqref{eq:Am1} in terms of $W'$-functions. It is straightforward to verify the final expression has all the right properties, such as obeying Ward identity for all the particles. 

The amplitude with one tachyon is closely related and is given as   
\begin{align}
  	&A_{n-1}(1\ldots n-2, (n-1)_{t})
    	=\left({1\over\alpha'}- p_{12\ldots n-2}^2 \right) \times \sum_{\Gamma_1,\Gamma_2} { (-1)^{n-3}W'(Z_{(1,\Gamma_1)},Z_{(\Gamma_{2},n-2)})\over d_{\Gamma_1}d_{\Gamma_2}}\, .
  \end{align} 
  
\section{Examples of $W'$-functions}\label{app3}

In this appendix we present further examples of $W'$-functions. More importantly, we will explain in some detail how they are determined by relabelling symmetry and physical properties  of amplitudes, such as factorizations. \\

\paragraph{Two gluons.}
For two gluons there are only two partitions 
\begin{align}
	\mathbf{P}_{12}(2)=(1,2)=W_0(12)\, , &&\mathbf{P}_{12}(1,1)=(1)|(2)=0\, .
\end{align}
Therefore in this case, we can only have, 
\begin{align}
	W'(12)=\frac{\alpha'}{1-\alpha' p_{12}^2} W_0(12)\, .
\end{align}
The denominator is the massive particle propagator in the $\rm DF^2+YM$ theory. As the maximal partition $\mathbf{P}_{12\ldots n}(1,1,\ldots,n)$ always has trivial contribution, we will ignore this partition in the following discussions.

\paragraph{Three gluons.}
For three gluons, the non-trivial partitions are 
\begin{align}
	\mathbf{P}_{123}(3)=(1,2,3)\, , &&\mathbf{P}_{123}(1,2)=(1)|(2,3)\, , &&\mathbf{P}_{123}(2,1)=(1,2)|(3)\, . 
\end{align}
It is easy to check that the following symmetry property, 
\begin{align}
	\mathbb{O}_{{\rm cyc}([12]3)}\circ \mathbf{P}_{123}(1,2)=\mathbb{O}_{{\rm cyc}([12]3)}\circ \mathbf{P}_{123}(2,1)\, .
\end{align}
Therefore, we only have two independent terms  
\begin{align}
    \mathbf{P}_{123}(1,2,3)\, , \qquad  \mathbb{O}_{{\rm cyc}([12]3)}\circ \mathbf{P}_{123}(1,2) \, .
\end{align}
Their coefficients are then fixed by  factorization of gluons, and we obtain
\begin{align}
	W'(123)=\frac{\alpha'}{1-\alpha' p_{123}^2} \Big(\mathbf{P}_{123}(1,2,3)+ \mathbb{O}_{{\rm cyc}([12]3)}\circ \mathbf{P}_{123}(1,2)\Big)\, ,
\end{align}
where, as in \eqref{eq:defP}, $\mathbf{P}_{123}(1,2,3)$ and $\mathbf{P}_{123}(1,2)$ can be expressed in terms physical kinematics variables, which are given as
\begin{align}
    \mathbf{P}_{123}(1,2,3)=W_0(123)\, , \qquad \mathbf{P}_{123}(1,2)=p_3\mdot F_1\mdot p_2 W'(23) \, .
\end{align}

\paragraph{Four gluons.}
For four gluons, all the partitions are 
\begin{align}
	\mathbf{P}_{1234}(4)=(1,2,3,4)\, , &&\mathbf{P}_{1234}(1,3)=(1)|(2,3,4)\,, &&\mathbf{P}_{1234}(3,1)=(1,2,3)|(4)\,,\nn\\
	\mathbf{P}_{1234}(1,1,2)=(1)|(2)|(3,4)\,, &&\mathbf{P}_{1234}(1,2,1)=(1)|(2,3)|(4)\,, \nn\\
	\mathbf{P}_{1234}(2,1,1)=(1,2)|(3)|(4)\,,&&
	\mathbf{P}_{1234}(2,2)=(1,2)|(3,4)\,.
\end{align}
It is easy to see the symmetry properties, such as  
\begin{align}
	\mathbb{O}_{{\rm cyc}([123]4)}\circ \mathbf{P}_{1234}(1,3)&=\mathbb{O}_{{\rm cyc}([123]4)}\circ \mathbf{P}_{1233}(3,1)\, ,\nn\\
	\mathbb{O}_{{\rm cyc}([123]4)}\circ \mathbf{P}_{1234}(1,1,2)&=\mathbb{O}_{{\rm cyc}([123]4)}\circ \mathbf{P}_{1233}(1,2,1)=\mathbb{O}_{{\rm cyc}([123]4)}\circ \mathbf{P}_{1233}(2,1,1) \, , 
\end{align}
and $\mathbb{O}_{{\rm cyc}([123]4)}\circ \mathbf{P}_{1234}(2,2)$ is self-invariant, so we should take into account the over-counting and divide by the symmetry factor, which is $2$ in this case. 

In this case, we have four independent terms, 
\begin{align}
    \mathbf{P}_{1234}(5)\, , \quad \mathbb{O}_{{\rm cyc}([123]4)}\circ \mathbf{P}_{1234}(1,3)\, , \quad \mathbb{O}_{{\rm cyc}([123]4)}\circ \mathbf{P}_{1234}(1,1,2)\, , \quad \mathbb{O}_{{\rm cyc}([123]4)}\circ{1\over 2}\mathbf{P}_{1234}(2,2) \, . 
\end{align}
We fix their coefficients again by physical requirements of factorization,  and we find, 
\begin{align}
W'(1234)=\frac{\alpha'}{1-\alpha' p_{1234}^2} \Big(\mathbf{P}_{1234}(5)+ \mathbb{O}_{{\rm cyc}([123]4)}\circ \Big(\mathbf{P}_{1234}(1,3)+\mathbf{P}_{1234}(1,1,2)+{1\over 2}\mathbf{P}_{1234}(2,2)\Big)\Big)\, .
 \end{align}
 In terms of physical kinematics variables, it can be written as
 \begin{align}
& \!\! W'(1234)=\frac{4 p_1\mdot F_3\mdot p_4 p_1\mdot F_4\mdot p_2 W_0(12) \alpha'^3}{\left(1-\alpha'  p_{12}^2\right) \left(1-\alpha'  p_{1234}^2\right) \left(1-\alpha'  p_{124}^2\right)} {+} \frac{4 p_1\mdot F_4\mdot p_2 p_2\mdot F_3\mdot p_4 W_0(12) \alpha'^3}{\left(1-\alpha'  p_{12}^2\right) \left(1-\alpha'  p_{1234}^2\right) \left(1-\alpha'  p_{124}^2\right)} {+} \frac{4 p_1\mdot F_2\mdot p_3 p_1\mdot F_4\mdot p_3 W_0(13) \alpha'^3}{\left(1-\alpha'  p_{123}^2\right) \left(1-\alpha'  p_{1234}^2\right) \left(1-\alpha'  p_{13}^2\right)}\nn\\
 &+\frac{4 p_1\mdot F_2\mdot p_3 p_2\mdot F_4\mdot p_3 W_0(13) \alpha'^3}{\left(1-\alpha'  p_{123}^2\right) \left(1-\alpha'  p_{1234}^2\right) \left(1-\alpha'  p_{13}^2\right)}+\frac{4 p_1\mdot F_2\mdot p_3 p_1\mdot F_4\mdot p_3 W_0(13) \alpha'^3}{\left(1-\alpha'  p_{1234}^2\right) \left(1-\alpha'  p_{13}^2\right) \left(1-\alpha'  p_{134}^2\right)}+\frac{4 p_1\mdot F_2\mdot p_4 p_1\mdot F_4\mdot p_3 W_0(13) \alpha'^3}{\left(1-\alpha'  p_{1234}^2\right) \left(1-\alpha'  p_{13}^2\right) \left(1-\alpha'  p_{134}^2\right)}\nn\\
 &+\frac{2 p_1\mdot p_2 p_3\mdot p_4 W_0(13) W_0(24) \alpha'^3}{\left(1-\alpha'  p_{1234}^2\right) \left(1-\alpha'  p_{13}^2\right) \left(1-\alpha'  p_{24}^2\right)}+\frac{4 p_1\mdot F_3\mdot p_4 p_2\mdot F_1\mdot p_4 W_0(24) \alpha'^3}{\left(1-\alpha'  p_{1234}^2\right) \left(1-\alpha'  p_{124}^2\right) \left(1-\alpha'  p_{24}^2\right)}+\frac{4 p_2\mdot F_1\mdot p_4 p_2\mdot F_3\mdot p_4 W_0(24) \alpha'^3}{\left(1-\alpha'  p_{1234}^2\right) \left(1-\alpha'  p_{124}^2\right) \left(1-\alpha'  p_{24}^2\right)}\nn\\
 &+\frac{4 p_2\mdot F_1\mdot p_3 p_2\mdot F_3\mdot p_4 W_0(24) \alpha'^3}{\left(1-\alpha'  p_{1234}^2\right) \left(1-\alpha'  p_{234}^2\right) \left(1-\alpha'  p_{24}^2\right)}+\frac{4 p_2\mdot F_1\mdot p_4 p_2\mdot F_3\mdot p_4 W_0(24) \alpha'^3}{\left(1-\alpha'  p_{1234}^2\right) \left(1-\alpha'  p_{234}^2\right) \left(1-\alpha'  p_{24}^2\right)}+\frac{2 p_1\mdot p_3 p_2\mdot p_4 W_0(12) W_0(34) \alpha'^3}{\left(1-\alpha'  p_{12}^2\right) \left(1-\alpha'  p_{1234}^2\right) \left(1-\alpha'  p_{34}^2\right)}\nn\\
 &+\frac{4 p_1\mdot F_2\mdot p_3 p_3\mdot F_1\mdot p_4 W_0(34) \alpha'^3}{\left(1-\alpha'  p_{1234}^2\right) \left(1-\alpha'  p_{134}^2\right) \left(1-\alpha'  p_{34}^2\right)}+\frac{4 p_1\mdot F_2\mdot p_4 p_3\mdot F_1\mdot p_4 W_0(34) \alpha'^3}{\left(1-\alpha'  p_{1234}^2\right) \left(1-\alpha'  p_{134}^2\right) \left(1-\alpha'  p_{34}^2\right)}-\frac{4 p_1\mdot F_3\mdot p_2 p_1\mdot F_4\mdot p_3 W_0(12) \alpha'^3}{\left(1-\alpha'  p_{12}^2\right) \left(1-\alpha'  p_{123}^2\right) \left(1-\alpha'  p_{1234}^2\right)}\nn\\
 &-\frac{4 p_1\mdot F_3\mdot p_2 p_2\mdot F_4\mdot p_3 W_0(12) \alpha'^3}{\left(1-\alpha'  p_{12}^2\right) \left(1-\alpha'  p_{123}^2\right) \left(1-\alpha'  p_{1234}^2\right)}-\frac{4 p_1\mdot F_2\mdot p_4 p_1\mdot F_3\mdot p_4 W_0(14) \alpha'^3}{\left(1-\alpha'  p_{1234}^2\right) \left(1-\alpha'  p_{124}^2\right) \left(1-\alpha'  p_{14}^2\right)}-\frac{4 p_1\mdot F_2\mdot p_4 p_2\mdot F_3\mdot p_4 W_0(14) \alpha'^3}{\left(1-\alpha'  p_{1234}^2\right) \left(1-\alpha'  p_{124}^2\right) \left(1-\alpha'  p_{14}^2\right)}\nn\\
 &-\frac{4 p_1\mdot F_2\mdot p_3 p_1\mdot F_3\mdot p_4 W_0(14) \alpha'^3}{\left(1-\alpha'  p_{1234}^2\right) \left(1-\alpha'  p_{134}^2\right) \left(1-\alpha'  p_{14}^2\right)}-\frac{4 p_1\mdot F_2\mdot p_4 p_1\mdot F_3\mdot p_4 W_0(14) \alpha'^3}{\left(1-\alpha'  p_{1234}^2\right) \left(1-\alpha'  p_{134}^2\right) \left(1-\alpha'  p_{14}^2\right)}-\frac{4 p_1\mdot F_4\mdot p_3 p_2\mdot F_1\mdot p_3 W_0(23) \alpha'^3}{\left(1-\alpha'  p_{123}^2\right) \left(1-\alpha'  p_{1234}^2\right) \left(1-\alpha'  p_{23}^2\right)}\nn\\
 &-\frac{4 p_2\mdot F_1\mdot p_3 p_2\mdot F_4\mdot p_3 W_0(23) \alpha'^3}{\left(1-\alpha'  p_{123}^2\right) \left(1-\alpha'  p_{1234}^2\right) \left(1-\alpha'  p_{23}^2\right)}-\frac{2 p_1\mdot p_2 p_3\mdot p_4 W_0(14) W_0(23) \alpha'^3}{\left(1-\alpha'  p_{1234}^2\right) \left(1-\alpha'  p_{14}^2\right) \left(1-\alpha'  p_{23}^2\right)}-\frac{4 p_2\mdot F_1\mdot p_3 p_2\mdot F_4\mdot p_3 W_0(23) \alpha'^3}{\left(1-\alpha'  p_{1234}^2\right) \left(1-\alpha'  p_{23}^2\right) \left(1-\alpha'  p_{234}^2\right)}\nn\\
 &-\frac{4 p_2\mdot F_1\mdot p_4 p_2\mdot F_4\mdot p_3 W_0(23) \alpha'^3}{\left(1-\alpha'  p_{1234}^2\right) \left(1-\alpha'  p_{23}^2\right) \left(1-\alpha'  p_{234}^2\right)}-\frac{2 p_1\mdot p_4 p_2\mdot p_3 W_0(12) W_0(34) \alpha'^3}{\left(1-\alpha'  p_{12}^2\right) \left(1-\alpha'  p_{1234}^2\right) \left(1-\alpha'  p_{34}^2\right)}-\frac{4 p_2\mdot F_1\mdot p_3 p_3\mdot F_2\mdot p_4 W_0(34) \alpha'^3}{\left(1-\alpha'  p_{1234}^2\right) \left(1-\alpha'  p_{234}^2\right) \left(1-\alpha'  p_{34}^2\right)}\nn\\
 &-\frac{4 p_2\mdot F_1\mdot p_4 p_3\mdot F_2\mdot p_4 W_0(34) \alpha'^3}{\left(1-\alpha'  p_{1234}^2\right) \left(1-\alpha'  p_{234}^2\right) \left(1-\alpha'  p_{34}^2\right)}+\frac{2 p_1\mdot F_3\mdot F_4\mdot p_2 W_0(12) \alpha'^2}{\left(1-\alpha'  p_{12}^2\right) \left(1-\alpha'  p_{1234}^2\right)}+\frac{2 p_1\mdot F_4\mdot p_3 W_0(123) \alpha'^2}{\left(1-\alpha'  p_{123}^2\right) \left(1-\alpha'  p_{1234}^2\right)}\nn\\
 &+\frac{2 p_2\mdot F_4\mdot p_3 W_0(123) \alpha'^2}{\left(1-\alpha'  p_{123}^2\right) \left(1-\alpha'  p_{1234}^2\right)}+\frac{2 p_1\mdot F_2\mdot F_4\mdot p_3 W_0(13) \alpha'^2}{\left(1-\alpha'  p_{1234}^2\right) \left(1-\alpha'  p_{13}^2\right)}+\frac{2 p_2\mdot F_1\mdot p_3 W_0(234) \alpha'^2}{\left(1-\alpha'  p_{1234}^2\right) \left(1-\alpha'  p_{234}^2\right)}+\frac{2 p_2\mdot F_1\mdot p_4 W_0(234) \alpha'^2}{\left(1-\alpha'  p_{1234}^2\right) \left(1-\alpha'  p_{234}^2\right)}\nn\\
 &+\frac{2 p_2\mdot F_1\mdot F_3\mdot p_4 W_0(24) \alpha'^2}{\left(1-\alpha'  p_{1234}^2\right) \left(1-\alpha'  p_{24}^2\right)}+\frac{2 p_3\mdot F_1\mdot F_2\mdot p_4 W_0(34) \alpha'^2}{\left(1-\alpha'  p_{1234}^2\right) \left(1-\alpha'  p_{34}^2\right)}-\frac{2 p_1\mdot F_4\mdot F_3\mdot p_2 W_0(12) \alpha'^2}{\left(1-\alpha'  p_{12}^2\right) \left(1-\alpha'  p_{1234}^2\right)}-\frac{2 p_1\mdot F_3\mdot p_4 W_0(124) \alpha'^2}{\left(1-\alpha'  p_{1234}^2\right) \left(1-\alpha'  p_{124}^2\right)}\nn\\
 &-\frac{2 p_2\mdot F_3\mdot p_4 W_0(124) \alpha'^2}{\left(1-\alpha'  p_{1234}^2\right) \left(1-\alpha'  p_{124}^2\right)}-\frac{2 p_1\mdot F_2\mdot p_3 W_0(134) \alpha'^2}{\left(1-\alpha'  p_{1234}^2\right) \left(1-\alpha'  p_{134}^2\right)}-\frac{2 p_1\mdot F_2\mdot p_4 W_0(134) \alpha'^2}{\left(1-\alpha'  p_{1234}^2\right) \left(1-\alpha'  p_{134}^2\right)}-\frac{2 p_1\mdot F_2\mdot F_3\mdot p_4 W_0(14) \alpha'^2}{\left(1-\alpha'  p_{1234}^2\right) \left(1-\alpha'  p_{14}^2\right)}\nn\\
 &-\frac{2 p_2\mdot F_1\mdot F_4\mdot p_3 W_0(23) \alpha'^2}{\left(1-\alpha'  p_{1234}^2\right) \left(1-\alpha'  p_{23}^2\right)}-\frac{2 p_3\mdot F_2\mdot F_1\mdot p_4 W_0(34) \alpha'^2}{\left(1-\alpha'  p_{1234}^2\right) \left(1-\alpha'  p_{34}^2\right)}-\frac{W_0(1234) \alpha' }{1-\alpha'  p_{1234}^2}\, .
\end{align}

\paragraph{Five gluons.}
We now consider the example of five gluons.
The single set partition is given by
\begin{align}
	\mathbf{P}_{12345}(5)\equiv W_0(12345)\, . 
\end{align}
The two sets partitions are  
\begin{align}
	\mathbf{P}_{12345}(1,4)&\equiv (1)|(2,3,4,5)\equiv W'(2345)p_5\mdot F_1\mdot p_2\, ,\nn\\
	\mathbf{P}_{12345}(2,3)&\equiv (1,2)|(3,4,5)\equiv W'(12)p_2\mdot p_3 W'(345) p_5\mdot p_1\, ,\nn\\
	\mathbf{P}_{12345}(3,2)&\equiv (1,2,3)|(4,5)\equiv W'(123)p_3\mdot p_4 W'(45) p_5\mdot p_1\, ,\nn\\
	\mathbf{P}_{12345}(4,1)&\equiv(1 ,2,3,4)|(5)\equiv W'(1234)p_4\mdot F_5\mdot p_1\, . 
\end{align}
Three and four sets partitions are similar, and they are listed below:  
\begin{align}
	&\mathbf{P}_{12345}(1,1,3)\,, ~~~ \mathbf{P}_{12345}(1,2,2)\,, ~~~  \mathbf{P}_{12345}(1,3,1)\,, ~~~  \mathbf{P}_{12345}(2,1,2)\,, ~~~  \mathbf{P}_{12345}(2,2,1)\,, ~~~  \mathbf{P}_{12345}(3,1,1)\, ,\nn\\
	&\mathbf{P}_{12345}(1,1,1,2) \,, ~~~ \mathbf{P}_{12345}(1,1,2,1)\,, ~~~ \mathbf{P}_{12345}(1,2,1,1)\,, ~~~ \mathbf{P}_{12345}(2,1,1,1)\, .
\end{align}
Below are the symmetry properties of the partitions at five points, 
\begin{align}
	\mathbb{O}_{{\rm cyc}([1234]5)}\circ  \mathbf{P}_{12345}(1,4)&=\mathbb{O}_{{\rm cyc}([1234]5)}\circ  \mathbf{P}_{12345}(4,1)\,,\nn\\
	\mathbb{O}_{{\rm cyc}([1234]5)}\circ  \mathbf{P}_{12345}(2,3)&=\mathbb{O}_{{\rm cyc}([1234]5)}\circ  \mathbf{P}_{12345}(3,2)\,,\nn\\
	\mathbb{O}_{{\rm cyc}([1234]5)}\circ  \mathbf{P}_{12345}(1,1,3)&=\mathbb{O}_{{\rm cyc}([1234]5)}\circ  \mathbf{P}_{12345}(1,3,1)=\mathbb{O}_{{\rm cyc}([1234]5)}\circ  \mathbf{P}_{12345}(3,1,1)\,,\nn\\
	\mathbb{O}_{{\rm cyc}([1234]5)}\circ  \mathbf{P}_{12345}(2,2,1)&=\mathbb{O}_{{\rm cyc}([1234]5)}\circ  \mathbf{P}_{12345}(1,2,2)=\mathbb{O}_{{\rm cyc}([1234]5)}\circ  \mathbf{P}_{12345}(2,1,2)\,,\nn\\
	\mathbb{O}_{{\rm cyc}([1234]5)}\circ  \mathbf{P}_{12345}(1,1,1,2)&=\mathbb{O}_{{\rm cyc}([1234]5)}\circ  \mathbf{P}_{12345}(1,1,2,1)=\mathbb{O}_{{\rm cyc}([1234]5)}\circ  \mathbf{P}_{12345}(1,2,1,1) \nn\\
	&=\mathbb{O}_{{\rm cyc}([1234]5)}\circ  \mathbf{P}_{12345}(2,1,1,1)\,. 
\end{align}

From the relabeling symmetry and gauge invariance, we find the function $W'(12345)$ should be a linear combination of 
\begin{align}
	\mathbf{P}_{12345}(5)\,, ~~~ \mathbf{P}_{12345}(1,4)\,, ~~~\mathbf{P}_{12345}(2,3)\,, ~~~\mathbf{P}_{12345}(1,1,3)\,, ~~~\mathbf{P}_{12345}(1,2,2)\,, ~~~\mathbf{P}_{12345}(1,1,1,2)\, .
\end{align}
Therefore, there are six free parameters to be fixed, which once again are determined by the factorization behaviour of gluons. It turns out they are all identical $1$, and we get 
\begin{align}
	&W'(12345)=\frac{\alpha'}{1-\alpha' p_{12345}^2}\mathbf{P}_{12345}(5)+\frac{\alpha'}{1-\alpha' p_{12345}^2}\mathbb{O}_{{\rm cyc}([1234]5)} \circ \nn\\
    & \Bigg[\mathbf{P}_{12345}(1,4)+\mathbf{P}_{12345}(2,3)+\mathbf{P}_{12345}(1,1,3)+\mathbf{P}_{12345}(1,2,2)+\mathbf{P}_{12345}(1,1,1,2))\Bigg].
\end{align}


\section{Local BCJ numerators in YM theory with $\alpha'$ corrections}\label{app4}

In this appendix, we provide further non-trivial examples for local BCJ numerators in YM theory with $\alpha'$ corrections. \\

\paragraph{Four point BCJ numerator at $\alpha'$ order} 
\begin{align}
	&\npre^{(1)}(1234)=W_0(12) p_2\mdot F_3\mdot \veps _4-W_0(12) \veps _4\mdot p_2 \veps _3\mdot p_1-W_0(12) \veps _3\mdot p_2 \veps _4\mdot p_2-W_0(13) \veps _2\mdot p_1 \veps _4\mdot p_3+\frac{1}{2} W_0(23) p_1\mdot p_2 \veps _1\mdot \veps _4\nn\\
	&+W_0(23) \veps _1\mdot p_2 \veps _4\mdot p_3+W_0(123) \veps _4\mdot p_3+2 \veps _1\mdot \veps _4 p_1\mdot F_2\mdot p_3 \veps _3\mdot p_1+4 p_1\mdot p_2 \veps _1\mdot F_2\mdot \veps _4 \veps _3\mdot p_1+2 p_1\mdot p_2 \veps _2\mdot F_1\mdot \veps _4 \veps _3\mdot p_1\nn\\
	&+2 \veps _3\mdot \veps _4 p_1\mdot F_2\mdot p_3 \veps _1\mdot p_3+4 p_1\mdot p_2 \veps _1\mdot F_2\mdot \veps _4 \veps _3\mdot p_2+4 p_1\mdot p_3 \veps _1\mdot F_3\mdot \veps _4 \veps _2\mdot p_1-4 p_1\mdot F_2\mdot p_3 \veps _1\mdot F_3\mdot \veps _4+2 p_1\mdot p_2 \veps _2\mdot F_1\mdot \veps _4 \veps _3\mdot p_2\nn\\
	&+2 p_1\mdot p_3 \veps _3\mdot F_1\mdot \veps _4 \veps _2\mdot p_1-2 p_1\mdot F_2\mdot p_3 \veps _3\mdot F_1\mdot \veps _4-2 p_1\mdot p_2 \veps _1\mdot \veps _4 \veps _2\mdot p_1 \veps _3\mdot p_1-2 p_1\mdot p_3 \veps _1\mdot \veps _4 \veps _2\mdot p_1 \veps _3\mdot p_1-2 p_1\mdot p_2 \veps _2\mdot \veps _4 \veps _1\mdot p_2 \veps _3\mdot p_1\nn\\
	&-2 p_1\mdot p_2 \veps _1\mdot \veps _4 \veps _2\mdot p_1 \veps _3\mdot p_2-2 p_1\mdot p_2 \veps _2\mdot \veps _4 \veps _1\mdot p_2 \veps _3\mdot p_2-2 p_1\mdot p_3 \veps _3\mdot \veps _4 \veps _1\mdot p_3 \veps _2\mdot p_1 \, .
\end{align}
\paragraph{Four point BCJ numerator at $\alpha'^2$ order} 
\begin{align}
	&\npre^{(2)}(1234)=2 W_0(12) p_1\mdot p_2 p_2\mdot F_3\mdot \veps _4-2 W_0(12) p_1\mdot F_3\mdot p_2 \veps _4\mdot p_3+2 W_0(13) p_1\mdot F_2\mdot p_3 \veps _4\mdot p_2-W_0(23) p_2\mdot F_1\mdot p_3 \veps _4\mdot p_2\nn\\
	&-W_0(23) p_2\mdot F_1\mdot p_3 \veps _4\mdot p_3-2 W_0(12) \left(p_1\mdot p_2\right){}^2 \veps _3\mdot \veps _4-W_0(23) \left(p_1\mdot p_2\right){}^2 \veps _1\mdot \veps _4-2 W_0(12) p_1\mdot p_2 \veps _3\mdot p_1 \veps _4\mdot p_3\nn\\
	&-2 W_0(12) p_1\mdot p_2 \veps _3\mdot p_2 \veps _4\mdot p_3+2 W_0(13) p_1\mdot p_3 p_1\mdot p_2 \veps _2\mdot \veps _4-W_0(23) p_1\mdot p_3 p_1\mdot p_2 \veps _1\mdot \veps _4-W_0(23) p_1\mdot p_2 \veps _1\mdot p_2 \veps _4\mdot p_2\nn\\
	&-W_0(23) p_1\mdot p_2 \veps _1\mdot p_2 \veps _4\mdot p_3-2 W_0(13) p_1\mdot p_3 \veps _2\mdot p_1 \veps _4\mdot p_2-W_0(23) p_1\mdot p_3 \veps _1\mdot p_2 \veps _4\mdot p_2-W_0(23) p_1\mdot p_3 \veps _1\mdot p_2 \veps _4\mdot p_3 \, .
\end{align}
\paragraph{Four point BCJ numerator at $\alpha'^3$ order} 
\begin{align}
	&\npre^{(3)}(1234)=-4 W_0(12) \left(p_1\mdot p_2\right){}^2 p_2\mdot F_3\mdot \veps _4-4 W_0(12) p_1\mdot p_2 p_1\mdot F_3\mdot p_2 \veps _4\mdot p_3+2 W_0(23) p_1\mdot p_2 p_2\mdot F_1\mdot p_3 \veps _4\mdot p_2\nn\\
	&+2 W_0(23) p_1\mdot p_2 p_2\mdot F_1\mdot p_3 \veps _4\mdot p_3+4 W_0(13) p_1\mdot p_3 p_1\mdot F_2\mdot p_3 \veps _4\mdot p_2+2 W_0(23) p_1\mdot p_3 p_2\mdot F_1\mdot p_3 \veps _4\mdot p_2\nn\\
	&+2 W_0(23) p_1\mdot p_3 p_2\mdot F_1\mdot p_3 \veps _4\mdot p_3-4 W_0(12) \left(p_1\mdot p_2\right){}^3 \veps _3\mdot \veps _4+2 W_0(23) \left(p_1\mdot p_2\right){}^3 \veps _1\mdot \veps _4-4 W_0(12) \left(p_1\mdot p_2\right){}^2 \veps _3\mdot p_1 \veps _4\mdot p_3\nn\\
	&-4 W_0(12) \left(p_1\mdot p_2\right){}^2 \veps _3\mdot p_2 \veps _4\mdot p_3+4 W_0(23) p_1\mdot p_3 \left(p_1\mdot p_2\right){}^2 \veps _1\mdot \veps _4+2 W_0(23) \left(p_1\mdot p_2\right){}^2 \veps _1\mdot p_2 \veps _4\mdot p_2\nn\\
	&+2 W_0(23) \left(p_1\mdot p_2\right){}^2 \veps _1\mdot p_2 \veps _4\mdot p_3+4 W_0(13) \left(p_1\mdot p_3\right){}^2 p_1\mdot p_2 \veps _2\mdot \veps _4+2 W_0(23) \left(p_1\mdot p_3\right){}^2 p_1\mdot p_2 \veps _1\mdot \veps _4\nn\\
	&+4 W_0(23) p_1\mdot p_3 p_1\mdot p_2 \veps _1\mdot p_2 \veps _4\mdot p_2+4 W_0(23) p_1\mdot p_3 p_1\mdot p_2 \veps _1\mdot p_2 \veps _4\mdot p_3-4 W_0(13) \left(p_1\mdot p_3\right){}^2 \veps _2\mdot p_1 \veps _4\mdot p_2\nn\\
	&+2 W_0(23) \left(p_1\mdot p_3\right){}^2 \veps _1\mdot p_2 \veps _4\mdot p_2+2 W_0(23) \left(p_1\mdot p_3\right){}^2 \veps _1\mdot p_2 \veps _4\mdot p_3 \, .
\end{align}
\paragraph{Five point local BCJ numerator of pure Yang-Mills theory} 
\begin{align}
	&\npre^{\rm (YM)}(12345)=\frac{1}{6} \veps _1\mdot \veps _5 \veps _2\mdot p_1 p_{12}\mdot \veps _3 p_{123}\mdot \veps _4-\frac{1}{3} \veps _1\mdot \veps _5 p_{12}\mdot \veps _3 p_1\mdot F_2\mdot \veps _4-\frac{1}{6} \veps _1\mdot \veps _5 \veps _2\mdot p_1 p_{12}\mdot F_3\mdot \veps _4-\frac{1}{2} \veps _2\mdot p_1 p_{12}\mdot \veps _3 \veps _1\mdot F_4\mdot \veps _5\nn\\
 &+\frac{1}{6} \veps _1\mdot \veps _5 \veps _2\mdot p_1 p_{12}\mdot \veps _3 p_{12}\mdot \veps _4+\frac{1}{6} \veps _1\mdot \veps _5 \veps _4\mdot p_1 \veps _2\mdot p_1 p_{12}\mdot \veps _3-\frac{1}{3} \veps _1\mdot \veps _5 p_{123}\mdot \veps _4 p_1\mdot F_2\mdot \veps _3-\frac{1}{2} \veps _2\mdot p_1 p_{123}\mdot \veps _4 \veps _1\mdot F_3\mdot \veps _5\nn\\
 &+\frac{1}{6} \veps _1\mdot \veps _5 \veps _3\mdot p_1 \veps _2\mdot p_1 p_{123}\mdot \veps _4-\frac{1}{2} \veps _2\mdot p_1 p_{13}\mdot \veps _4 \veps _1\mdot F_3\mdot \veps _5+\frac{1}{6} \veps _1\mdot \veps _5 \veps _3\mdot p_1 \veps _2\mdot p_1 p_{13}\mdot \veps _4+\frac{1}{2} \veps _1\mdot F_3\mdot \veps _5 p_1\mdot F_2\mdot \veps _4+\frac{1}{2} \veps _1\mdot F_4\mdot \veps _5 p_1\mdot F_2\mdot \veps _3\nn\\
 &-\frac{1}{6} \veps _1\mdot \veps _5 \veps _4\mdot p_1 p_1\mdot F_2\mdot \veps _3-\frac{1}{6} \veps _1\mdot \veps _5 \veps _3\mdot p_1 p_1\mdot F_2\mdot \veps _4+\frac{1}{3} \veps _1\mdot \veps _5 p_1\mdot F_{23}\mdot \veps _4+\veps _2\mdot p_1 \veps _1\mdot F_3\mdot F_4\mdot \veps _5-\frac{1}{3} \veps _1\mdot \veps _5 \veps _2\mdot p_1 p_1\mdot F_3\mdot \veps _4\nn\\
 &-\frac{1}{2} \veps _3\mdot p_1 \veps _2\mdot p_1 \veps _1\mdot F_4\mdot \veps _5+\frac{1}{6} \veps _1\mdot \veps _5 \veps _3\mdot p_1 \veps _4\mdot p_1 \veps _2\mdot p_1-\frac{1}{2} p_{12}\mdot \veps _3 p_{123}\mdot \veps _4 \veps _1\mdot F_2\mdot \veps _5+\frac{1}{2} \veps _1\mdot F_2\mdot \veps _5 p_{12}\mdot F_3\mdot \veps _4+p_{12}\mdot \veps _3 \veps _1\mdot F_2\mdot F_4\mdot \veps _5\nn\\
 &-\frac{1}{2} p_{12}\mdot \veps _3 p_{12}\mdot \veps _4 \veps _1\mdot F_2\mdot \veps _5+p_{123}\mdot \veps _4 \veps _1\mdot F_2\mdot F_3\mdot \veps _5-\veps _1\mdot F_2\mdot F_3\mdot F_4\mdot \veps _5 \, .
\end{align}
Other five point BCJ numerators in YM theory in the DDM basis are obtained by the crossing permutation over the leg $2,3,4$. 
More examples of BCJ numerators and a {\tt Mathematica} package can be found at  {\href{https://github.com/AmplitudeGravity/kinematicHopfAlgebra}{{\it \blue KiHA5.0} GitHub repository}}~\cite{ChenGitHub}.

\end{document}